\documentclass[aps, pre, 12pt]{revtex4-1}
\usepackage{amsthm}
\usepackage{subfigure}
\pdfoutput=1
\usepackage{graphicx}
\usepackage{epsfig}
\usepackage{color}
\usepackage{amsmath}
\usepackage{amssymb}
\usepackage{ulem}
\begin{document}

\title{Effect of P\'{e}clet number on miscible rectilinear displacement in a Hele-Shaw cell}

\author{Satyajit Pramanik, Manoranjan Mishra}
\affiliation{Department of Mathematics, Indian Institute of Technology Ropar,  Rupnagar - 140001, Punjab, India}

\date{\today}

\begin{abstract}
Influence of fluid dispersion on the Saffman-Taylor instability in miscible fluids has been investigated both in the linear and nonlinear regimes. The convective characteristic scales are used for the dimensionless formulation that incorporates P\'{e}clet number (Pe) into the governing equations as a measure for the fluid dispersion. A linear stability analysis (LSA) has been performed in a similarity transformation domain using the quasi-steady-state approximation. LSA results show that systems with large Pe become more unstable and the onset of instability occurs earlier compared to the case when Pe is smaller. Variations of the most unstable wave number and the cut-off wave number with Pe have been analyzed. Fourier spectral method has been used for the numerical simulations of the fully nonlinear system. The results indicate that the wave numbers of the unstable modes increase with Pe. Influence of the anisotropic dispersion on the onset in both the linear and nonlinear regimes has been analyzed. Large transverse diffusivity increases the length scale of the fingers quickly and merges the fingers to generate coarser fingers. Finally the combined effect of the Korteweg stress and Pe in the linear regime has been perused. In the presence of the Korteweg stresses and depending upon various flow parameters, a fluid system with larger Pe exhibits smaller instantaneous growth rate than with smaller Pe, which is a counter-intuitive result. 
\end{abstract}

\maketitle

\section{Introduction}\label{sec:introduction}
Displacement mechanism in porous media plays significant role in several industrial and environmental processes, such as secondary and tertiary oil recovery \cite{H}, hydrology \cite{DBM}, chromatography \cite{RDM, MMD, DBM}, aquifers \cite{DBM} and geodynamics \cite{MY}, etc. Recently, over the last one decades and half researchers and scientists face challenges to capture the environmental $\text{CO}_2$ and store them for longer period in order to reduce the amount of carbon released in the air. This also involves the displacement in porous rock. Looking into the vast area of industrial and real life applications it is very important to study such multiphase flow problems in porous media. In all such displacement processes a convective instability takes place when a less viscous fluid displaces another fluid of higher viscosity. These instability patterns looks like fingers - hence called viscous fingering (VF) instability or the Saffman-Taylor instability \cite{ST} (sometimes also called Hill's instability \cite{HL}). Researchers have investigated about the dynamics of these instabilities over several decades for both the immiscible and miscible fluids. In the case of the former surface tension acts as a stabilization factor, whereas in the miscible fluid dispersion plays an important role in the stabilization of the system. Motivated by the pioneering work of Saffman and Taylor \cite{ST} several authors carried out various LSA \cite{TH1, PM1, KC, TH3, YY} to capture the onset of instability and also performed nonlinear simulations \cite{TH2, CWM, ZH} to analyze the dynamic pattern of the instability. Tan and Homsy \cite{TH1} presented LSA for VF instability in classical single interface fluid displacement using quasi-steady-state-approximation (QSSA) method and initial value problem (IVP) approach. Following the work of Tan and Homsy \cite{TH1}, Kim and Choi \cite{KC} performed LSA using  QSSA in a similarity transformation domain and a spectral analysis. It has been shown that the results obtained from their analyses are physically more relevant than those of QSSA, and are also in accordance with the IVP results. Pramanik and Mishra \cite{PM1} carried out self-similar QSSA (SS-QSSA) to investigate the onset of instability for the single interface VF as well as displacement of localized fluid of finite extent both in the presence and absence of the Korteweg stresses. Their analyses determine the effect of width of the finite fluid on the onset of VF instability. 

Dimensionless formulation of the mathematical model is an integral and indispensable part of the theoretical investigations. In principle, such formulations are not unique and they depend upon the choice of the characteristic scales based on their presence in the model and/or the interest of the particular study. Two different types of characteristic scales are mainly used in the literature of miscible viscous fingering instability. One uses the diffusive length $\displaystyle\left(\frac{D_x}{U}\right)$ and time $\displaystyle\left(\frac{D_x}{U^2}\right)$ as the characteristic length and time scales,\cite{DBM, MMD, RDM, TH1, TH2, PM1, KC} whereas in the other convective length ($L_y$) and time $\displaystyle\left(\frac{L_y}{U}\right)$ \cite{TH3, YY, ZH, CWM} are used for the same. Here, $D_x$ and $U$ represent the axial dispersion coefficient and the uniform displacement speed of the fluid in a Hele-Shaw cell of width $L_y$. The latter formulation induces explicit dependence of the fluid dispersion in terms of the dimensionless parameter P\'{e}clet number, Pe, which is the ratio of the convective to the diffusive fluxes caused by the compositional gradients in the fluid. However, in the former explicit influence of fluid dispersion on miscible viscous fingering is hard to incorporate through linear stability analysis. Tan and Homsy \cite{TH3} performed a linear stability analysis to analyze the influence of Pe on the miscible radial source flow displacement in porous media. This study shows the existence of a critical P\'{e}clet number, Pe$_c$, below which diffusion is large enough such that all the perturbations remain stable. Zimmerman and Homsy \cite{ZH} carried out nonlinear simulations using the characteristic length of the model to investigate the influence of anisotropic dispersion in miscible VF. Their investigation elucidate that for large values of Pe one dimensional averaged concentration profile at large time remains invariant with Pe and anisotropic dispersion. Yang and Yortsos \cite{YY} presented an asymptotic analysis to understand the effect of Pe on miscible VF in Stokes flow regime between two parallel plates or in a cylindrical capillary. They determined a sharp upper bound for the Pe number below which the transverse averaging and the conventional convection-diffusion equation associated with the miscible Hele-Shaw flow is valid. 

Discussion of the above-mentioned literature reveals the fact that rigorous investigation about explicit dependence of fluid dispersion or Pe number on the miscible displacement is important in the context of VF in a Hele-Shaw cell. As mentioned earlier this aspect of miscible VF has been addressed by some of the researchers in the context of linear stability analysis for radial source flow \cite{TH3}, nonlinear simulations for rectilinear displacement of flat interface \cite{ZH} and localized circular fluid \cite{CWM}. However, to the best of the authors' knowledge, investigation about the effect of Pe on miscible rectilinear displacement in the linear regime was not attempted before. We follow a dimensionless formulation taking the convective characteristic length and time scales that incorporates Pe as the reciprocal of the dimensionless axial dispersion. The aim of the present study is to investigate the influence of fluid dispersion on the miscible VF instability through self-similar linear stability analysis \cite{PM1}. Next, nonlinear simulations have been performed using Fourier spectral method \cite{DBM, MMD, TH2} to investigate the possible effect of fluid dispersion on the interaction of fingers in the nonlinear regime. This is followed by comparison of the linear regime results of the rectilinear displacements with those of the radial source flow and the fully nonlinear simulations. 

It is clear that a slow diffusion mechanism gives rise to steep concentration gradient, which leads to intense fingering instability as diffusion acts as a stabilization factor in miscible fluid systems. However, it is well established both theoretically \cite{CWM, J, PM1} as well as experimentally \cite{MFMG_2007} that in the presence of steep concentration gradient a non-conventional stress, called the Korteweg stress \cite{K}, appears in miscible fluids. These gradient stresses stabilize the VF instability. Thus, two opposite phenomena take place out of the same physical consequence of steep concentration gradient that may lead to interplay between these two effects. This gives rise to another important question about miscible VF instability, how a destabilizing and a stabilizing force resulting from slow diffusion co-exist during the displacement of miscible fluids in a Hele-Shaw cell? These fundamental questions are addressed in the present study and we try to present a compact qualitative and quantitative analyses in this paper. 

The paper is organized in the following manner. The mathematical formulation of the problem is presented in Sec. \ref{sec:MF}. Sec. \ref{sec:LSA} discusses the linear stability analysis in a self-similar domain followed by the fully nonlinear simulations in Sec. \ref{sec:NS}. Next, we discuss the influence of Pe in the presence of the Korteweg stress in Sec. \ref{sec:IPeKS} followed by the concluding remarks in Sec. \ref{sec:C}. 

% figure1
\begin{figure}[ht]
\centering
\includegraphics[width=6.5in, keepaspectratio=true, angle=0]{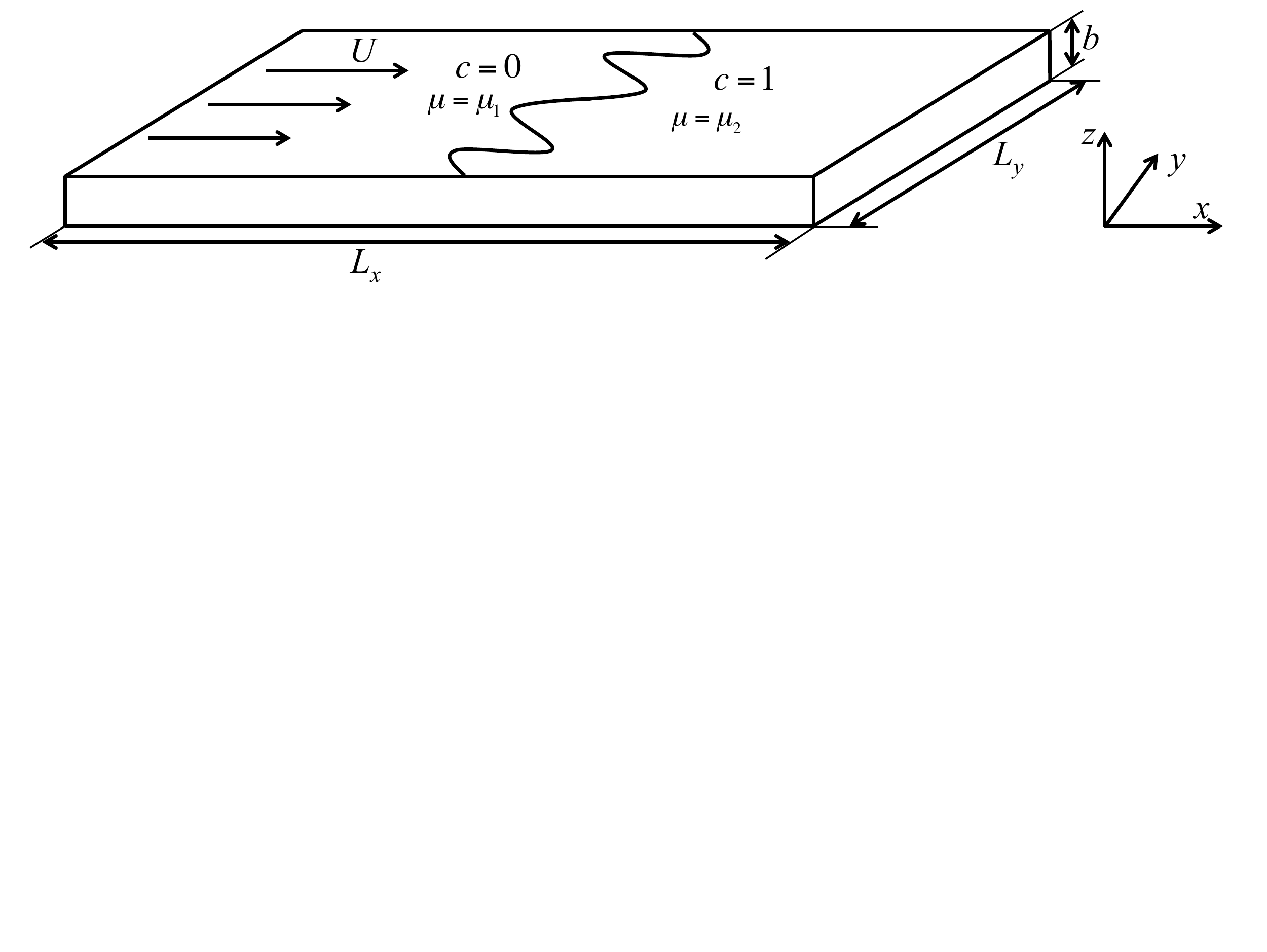}
\vspace{-3.8in}
\caption{Schematic diagram for the displacement of a miscible fluid by another fluid in a Hele-Shaw cell or 2D homogeneous porous media.}
\label{fig:schematic}
\end{figure}

\section{Mathematical formulation}\label{sec:MF}
\subsection{Governing equations}\label{subsec:GE}
Consider the displacement of two incompressible, neutrally buoyant fluids, which are miscible in all proportions in a Hele-Shaw cell or 2D homogeneous porous media as shown in Fig. \ref{fig:schematic}. Conservation of momentum in a Hele-Shaw cell is described by  Darcy's law, which is mathematically analogous to the flow in homogeneous porous media. The viscosities of the resident and the displacing fluids are $\mu_2$ and $\mu_1$, respectively, and they strongly depend on the solute concentration $c$. As mentioned earlier, we choose the width ($L_y$) of the Hele-Shaw cell as the characteristic length scale. The gap between the two parallel plates of the Hele-Shaw cell $b~ (\ll L_y)$ could be a possible length scale, \cite{Fernandez2001} which is equivalent to the square root of the permeability $\sqrt{\kappa}$ for a homogeneous porous media of permeability $\kappa = b^2/12$. As explained by Tan and Homsy \cite{TH3}, $\sqrt{\kappa}$ is not a suitable choice for the characteristic length scale, since Darcy's law is based upon the continuum hypothesis, we choose $L_y$ as the characteristic length scale in this paper. The characteristic time, velocity, pressure and viscosity are taken as $\displaystyle\frac{L_y}{U}, U, \displaystyle\frac{\mu_1UL_y}{\kappa}$ and $\mu_1$, respectively. Thus, the dimensionless equations governing such flow in a reference frame moving with the speed $U$ of the displacing fluid are, 
\begin{eqnarray}
\label{eq:dimcon}
& & \nabla\cdot\vec{u} = 0 \\
\label{eq:dimdarKorteweg}
& & \nabla p = -\mu (\vec{u} + \hat{i})\\
\label{eq:dimcondiff}
& & \frac{\partial c}{\partial t} + \vec{u}\cdot\nabla c = \frac{1}{\text{Pe}}\bigg(\frac{\partial^2c}{\partial x^2} + \epsilon \frac{\partial^2c}{\partial y^2}\bigg).
\end{eqnarray}

The dimensionless parameters appearing in the the equations are P\'{e}clet number and the anisotropic dispersion coefficient, defined as $\text{Pe} = \displaystyle\frac{UL_y}{D_x}$ and $\epsilon = \displaystyle\frac{D_y}{D_x}$, respectively. Here $D_y$ corresponds to the dispersion coefficient in the transverse direction, which is same as the molecular diffusion coefficient and the axial dispersion coefficient is $D_x = \displaystyle\frac{U^2b^2}{210D_y}$. Viscosity of the fluids are taken to be exponential function of the concentration $c$, i.e. $\mu(c) = \displaystyle{e^{Rc}}$, where $R$ is the logarithm of the ratio of the viscosities for the displaced to the displacing fluid. The fluid velocity is governed by the Darcy's law, Eq. \eqref{eq:dimdarKorteweg}, and the evolution of the solute concentration is measured through a convection-diffusion equation (Eq. \eqref{eq:dimcondiff}). Here $p$ is the hydrodynamic pressure and the velocity $\vec{u}$ has the stream-wise and span-wise components $u$ and $v$, respectively. These equations are solved simultaneously associated with appropriate boundary and initial conditions mentioned below.

\subsection{Boundary and initial conditions}\label{subsec:BIC}
Velocity of the fluid system is set in such a way that initially the complete system was moving with a uniform velocity $U$ along the stream-wise direction. The displacing fluid is kept on entering into the domain with that uniform speed resulting the resident fluid to exit at the same speed from the other end of the Hele-Shaw cell. Solute concentration does not experience any change across all the boundaries. Therefore, these can be written in the moving reference frame as,
\begin{eqnarray}
\label{eq:bc1}
& & \vec{u} = (0,0), ~~~ \frac{\partial c}{\partial x} = 0 ~~~ \text{as} ~~~ x \to \pm \infty, \\
\label{eq:bc2}
& & \frac{\partial v}{\partial y} = 0, ~~~ \frac{\partial c}{\partial y} = 0 ~~~ \text{for} ~~~ y = 0, 1.
\end{eqnarray} 

The velocity boundary conditions given in Eq. \eqref{eq:bc2} correspond to the constant pressure at the transverse boundaries \cite{NB}, where the stream-wise velocity $u$ is arbitrary. The initial distribution of the solute concentration is $c = 0$ and $c = 1$ in the displacing and displaced fluids, respectively. Hence, the initial condition can be written in the moving reference frame as,
\begin{eqnarray}
\label{eq:initial}
\vec{u} = (0,0) ~~~ \text{and} ~~~ c = \left\{\begin{matrix}
0, & x < 0 \\
1, & x \geq 0
\end{matrix}
\right., ~~~ \text{at} ~~~ t = 0.
\end{eqnarray} 

\section{Linear stability analysis}\label{sec:LSA}
Here we present a linear stability analysis to investigate the influence of Pe on the onset of VF instability following the SS-QSSA method \cite{PM1}. The base-state velocity field is assumed to be a null vector and the base-state concentration field is obtained by solving the 1D diffusion equation associated with no flux boundary conditions and the initial condition for the concentration given in Eq. \eqref{eq:initial}. Hence, the base flow can be written as,
\begin{eqnarray}
& & \vec{u}_0 = (0,0), \\
\label{eq:basecon}
& & c_0(\xi) = \frac{1}{2}\bigg(1 + \text{erf}\left(\frac{\xi}{2/\sqrt{\text{Pe}}}\right)\bigg), \\
\label{eq:basevisco}
& & \mu_0(\xi) = \mu_0(c_0),
\end{eqnarray}
where $\xi = \displaystyle\frac{x}{\sqrt{t}}$ is the similarity variable. Base-state concentration $c_0$ describes the self-similar diffusive decay of a step profile. An infinitesimal perturbation has been introduced in order to investigate the stability of the base state flow in the linear regime. Linearize the governing equations at a frozen diffusive time $t_0$ about the base flow to obtain the following equations in terms of the perturbation quantities,
\begin{eqnarray}
\label{eq:lincon}
& & \frac{1}{\sqrt{t_0}}\frac{\partial u'}{\partial \xi} + \frac{\partial v'}{\partial y} = 0, \\
\label{eq:lindarx}
& & \frac{1}{\sqrt{t_0}}\frac{\partial P'}{\partial \xi} = -\mu_0u' - \mu', \\
\label{eq:lindary}
& & \frac{\partial P'}{\partial y} = -\mu_0 v',\\
\label{eq:lincondiff}
& & \frac{\partial c'}{\partial t} - \frac{\xi}{2t_0}\frac{\partial c'}{\partial \xi} + \frac{1}{\sqrt{t_0}}\frac{\text{d}c_0}{\text{d}\xi}u' = \frac{1}{\text{Pe}}\bigg(\frac{1}{t_0}\frac{\partial^2c'}{\partial \xi^2} + \epsilon\frac{\partial^2c'}{\partial y^2}\bigg), \\
\label{eq:linvisco}
& & \mu' = \bigg(\frac{\text{d}\mu}{\text{d}c}\bigg|_{c_0}\bigg)c',
\end{eqnarray} 
in the similarity transformation domain. Eliminate the pressure and transverse velocity component from Eqs. \eqref{eq:lincon} - \eqref{eq:linvisco} and decompose the perturbation quantities $c', u'$  in the Fourier modes, \cite{PM1} $ (u',c')(\xi, y, t) = (\phi(\xi), \psi(\xi))\displaystyle{e^{iky + \sigma^*(k,t_0) t}} $. Here $k$ is the wave number of the sinusoidal perturbation and $\sigma^*$ is the instantaneous growth rate of the corresponding mode. Application of SS-QSSA method followed by simple algebraic calculations yield the following eigenvalue problem,
\begin{eqnarray}
\label{eq:EIG1}
& & \left(\frac{d^2}{d\xi^2} + R\frac{dc_0}{d\xi}\frac{d}{d\xi} - k^2t_0\right)\phi(\xi) = Rk^2t_0\psi(\xi), \\
\label{eq:EIG2}
& & \left(\sigma^*(k,t_0) - \frac{1}{\text{Pe}}\left(\frac{1}{t_0}\frac{d^2}{d\xi^2} - \epsilon k^2\right) - \frac{\xi}{2t_0}\frac{d}{d\xi}\right)\psi(\xi) = -\frac{1}{\sqrt{t_0}}\frac{dc_0}{d\xi}\phi(\xi).
\end{eqnarray}
Eigenvalues of the system of Eqs. \eqref{eq:EIG1} - \eqref{eq:EIG2} correspond the instantaneous growth rates of the perturbations.

Analytic solutions of the eigenvalue problem, Eqs. \eqref{eq:EIG1} - \eqref{eq:EIG2}, are not attainable. Hence it has been solved following the numerical technique described by Pramanik and Mishra \cite{PM1}. The obtained results are discussed in Sec. \ref{subsec:NR} to analyze the influence of fluid dispersion on VF instability in the linear regime away from the initial time, i.e. at $t_0 \neq 0$. Sec. \ref{subsec:CRSF} compares the results obtained with the radial source flow.	 It is worthy to mention that the dimensionless length and time in the two formulations discusses above differ by a factor of Pe. Putting Pe $= 1$ along with the isotropic dispersion ($\epsilon = 1$) in the present dimensionless formulation we retrive the analyses of Pramanik and Mishra \cite{PM1} in the absence of the Korteweg stresses. 

% figure2
\begin{figure}[ht]
(a) \hspace{8 cm} (b) \\
\centering
\includegraphics[width=3.2in, keepaspectratio=true, angle=0]{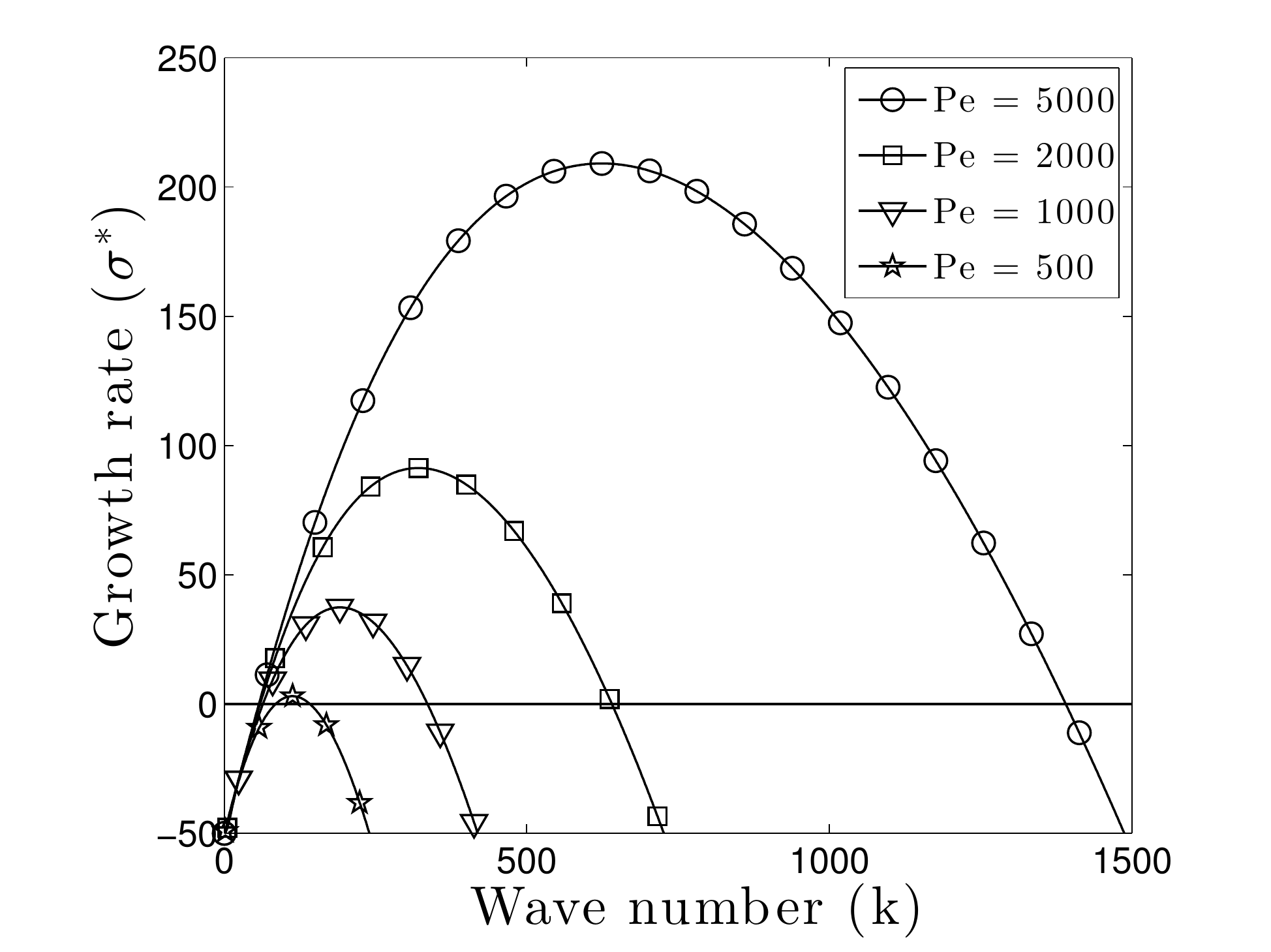}
\includegraphics[width=3.2in, keepaspectratio=true, angle=0]{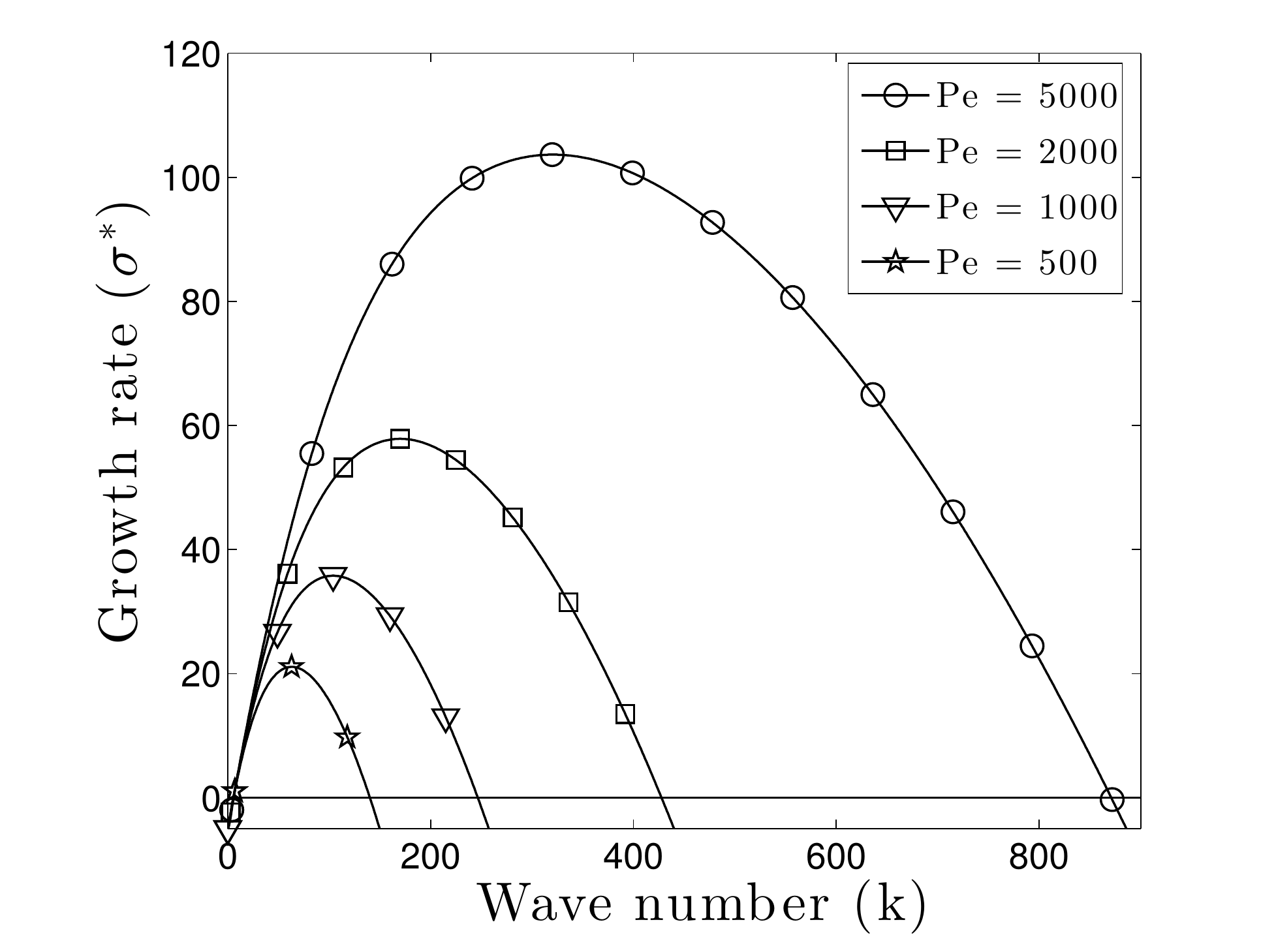}
\caption{Dispersion curves for $R = 3, \epsilon = 1$ with different Pe values at frozen diffusive time (a) $t_0 = 0.01$, (b) $t_0 = 0.1$. }
\label{fig:dispersion}
\end{figure}

\subsection{Numerical results}\label{subsec:NR}
The main aim of the present study is to investigate the influence of Pe number on VF instability. Fluid dispersion acts as a stabilization factor in miscible VF. Hence it is expected that instability will be more dominant when fluid diffuses at a slower rate, i.e. for higher Pe values. Firstly, we will analyze the influence of Pe number for the simpler case of isotropic dispersion, i.e. $\epsilon = 1$. Fig. \ref{fig:dispersion} shows the dispersion curves at two different frozen diffusive times $t_0 = 0.01, 0.1$ for different values of Pe with $R = 3$. It depicts that for both $t_0 = 0.01$ and $0.1$, the dispersion curve for the largest Pe lies above rest of the curves, which are aligned in the increasing order of Pe. Fig. \ref{fig:dispersion}(a) elucidates that for $\text{Pe} < 500$ all the perturbations have stable instantaneous growth rate. Thus there exists a critical value of Pe, $\text{Pe}_c$, below which dispersion is large enough to make all the perturbations stable. However, at $t_0 = 0.1$ the $\text{Pe}_c$ is not the same as that at $t_0 = 0.01$. It is clear that with a decrease in the fluid dispersion the most dangerous mode, \cite{PM1} $(k_{max}, \sigma^*_{max})$, corresponding the wave number of that particular disturbance having the largest instantaneous growth rate, increases. It is also noticed that the cutoff wave number,\cite{PM1} $k_c$, increases with Pe; however, the threshold wave number,\cite{PM1} $k_t$, remains almost unaltered with Pe. These signify that as the fluid dispersion becomes smaller the growth of the disturbances become larger and simultaneously the width of the intervals of unstable wave numbers increase. 

%figure3
\begin{figure}[ht]
\centering
\includegraphics[width=5in, keepaspectratio=true, angle=0]{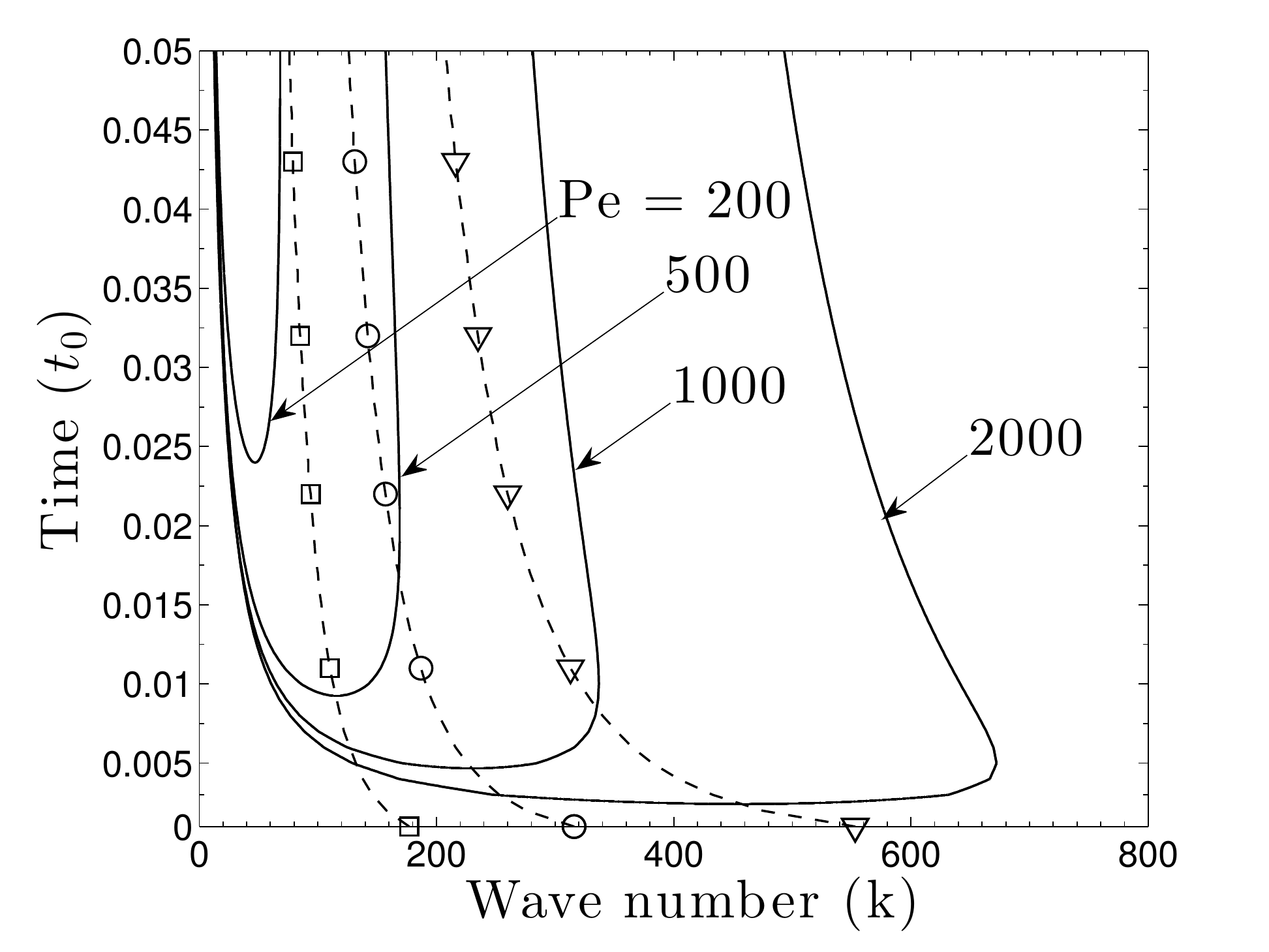}
\caption{Neutral stability curves ($\sigma^* = 0$) for different Pe values with $R = 3, \epsilon = 1$. Lines with markers correspond the locus of the $k_{max}$ at successive times: Pe $= 500 (\Box)$, Pe $= 1000 (\bigcirc)$, Pe $= 2000 (\bigtriangledown)$.}\label{fig:neutral}
\end{figure}

In order to identify the region of instability the neutral stability curves, contours of $\sigma^* = 0$, have been drawn for different Pe values on the $(k,t_0)$-plane. Fig. \ref{fig:neutral} depicts neutral stability curves for Pe = $500, 1000, 2000$ with $R = 3$. Area above the neutral stability curve represents the region of instability and the minimum point on this curve corresponds the critical point ($k_{cric}, t_{cric}$). This critical point represents the minimum pair of wave number, $k$, and the frozen time, $t_0$, for the specific values of $R$ and Pe, in order to the occurrence of instability. Fig. \ref{fig:neutral} clearly shows that the regions of instability increase with the increase of Pe, signifying that the slowest dispersion leads to the sever most instability. The dashed lines with markers represent the locus of the most unstable wave numbers $k_{max}$ at successive times $t_0$ and it is observed that these $k_{max}$ decrease rapidly at an early time and after some initial time they evolve very slowly. During this period of time for slow evolution of $k_{max}$, all the curves corresponding to different Pe values propagate almost parallel. 

%figure4
\begin{figure}[ht]
(a) \hspace{8 cm} (b) \\
\centering
\includegraphics[width=3.2in, keepaspectratio=true, angle=0]{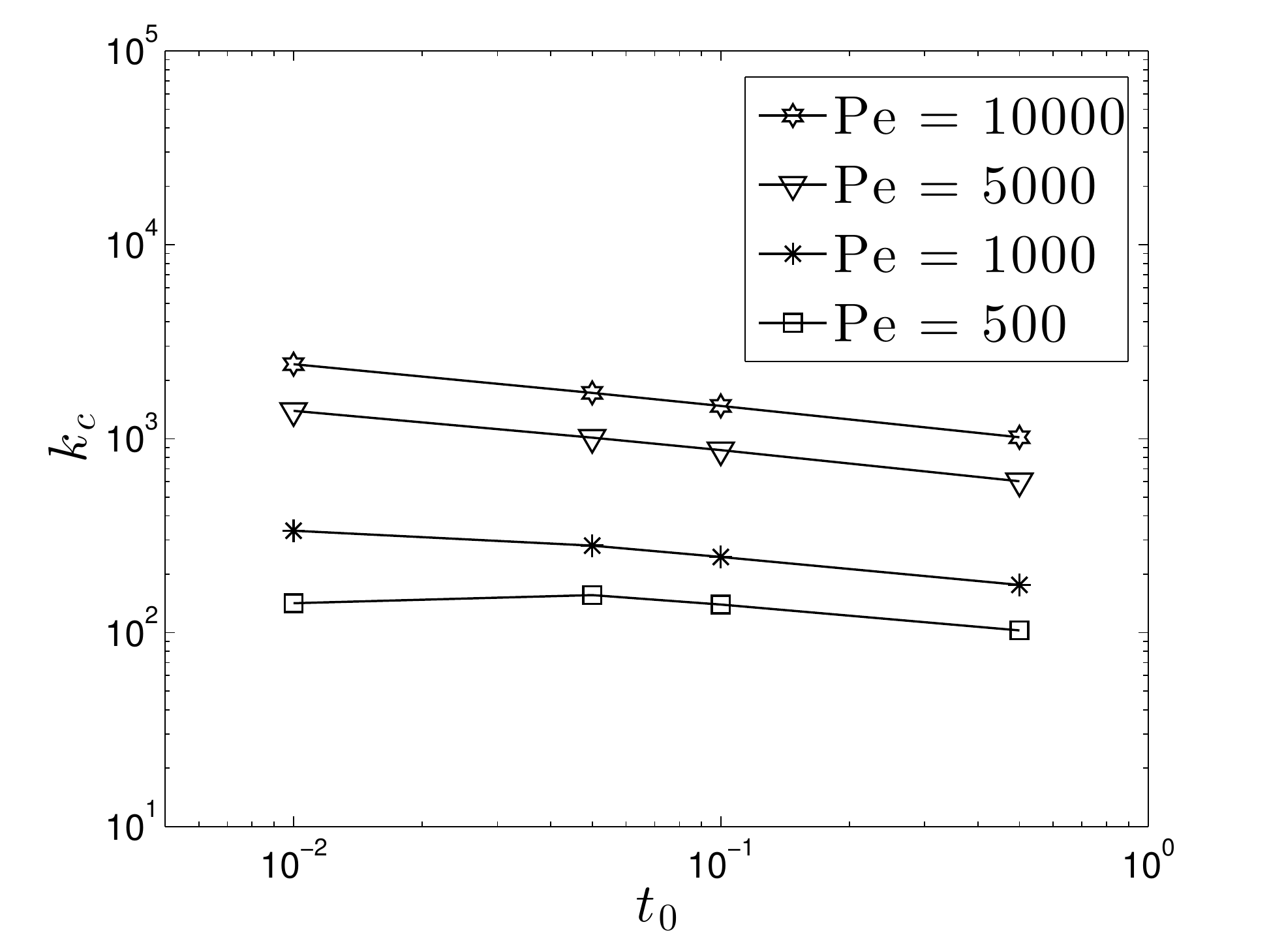}
\includegraphics[width=3.2in, keepaspectratio=true, angle=0]{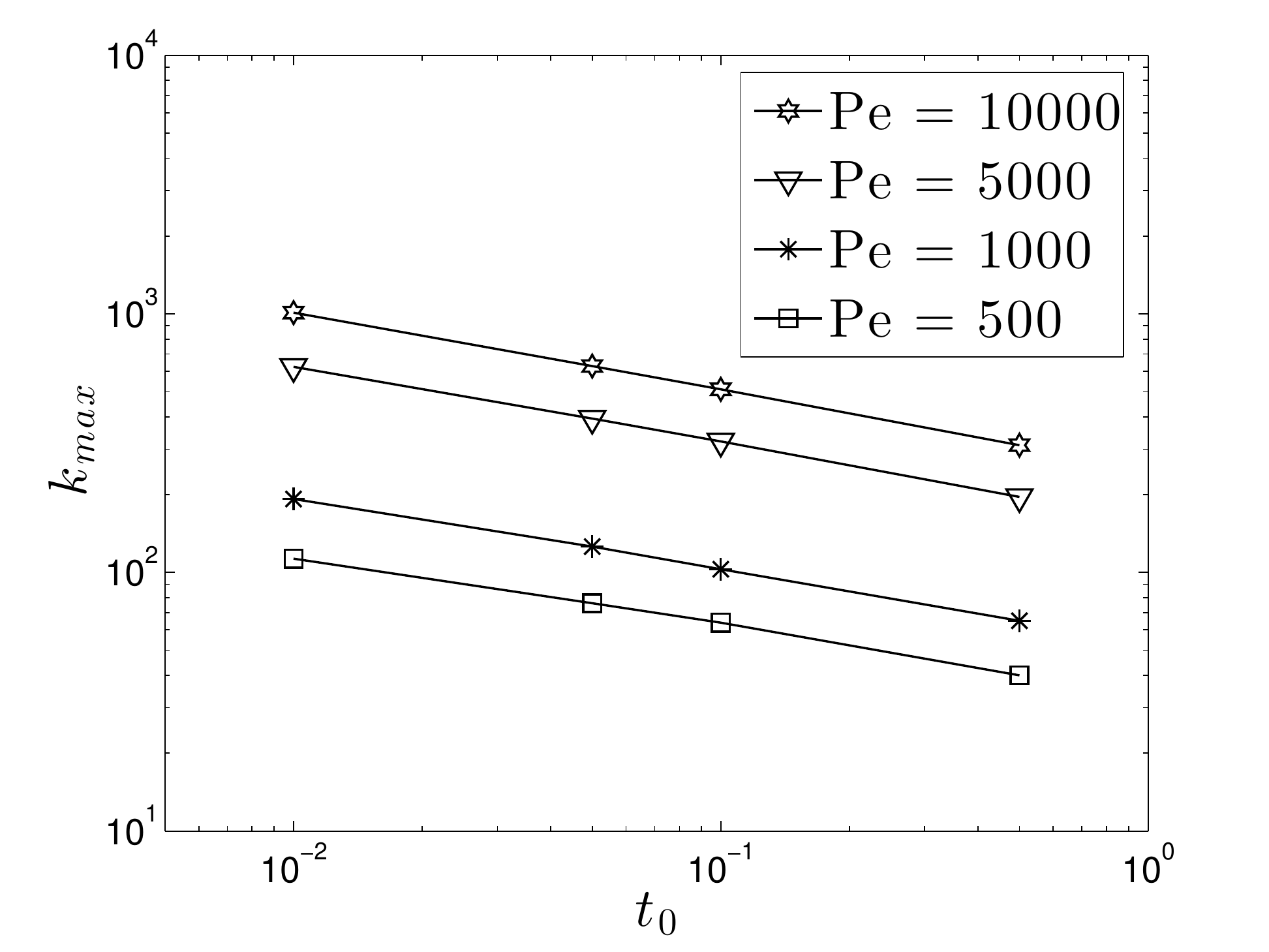}
\caption{(a) Cutoff wave number $k_c$ and (b) the most dangerous wave number $k_{max}$ as a function of frozen diffusive time $t_0$ for $R = 3$; $k_c \sim t_0^{-0.22}$ and  $k_{max} \sim t_0^{-0.3}$. }\label{fig:cutofft0}
\end{figure}

%figure5
\begin{figure}[ht]
(a) \hspace{8 cm} (b) \\
\centering
\includegraphics[width=3.2in, keepaspectratio=true, angle=0]{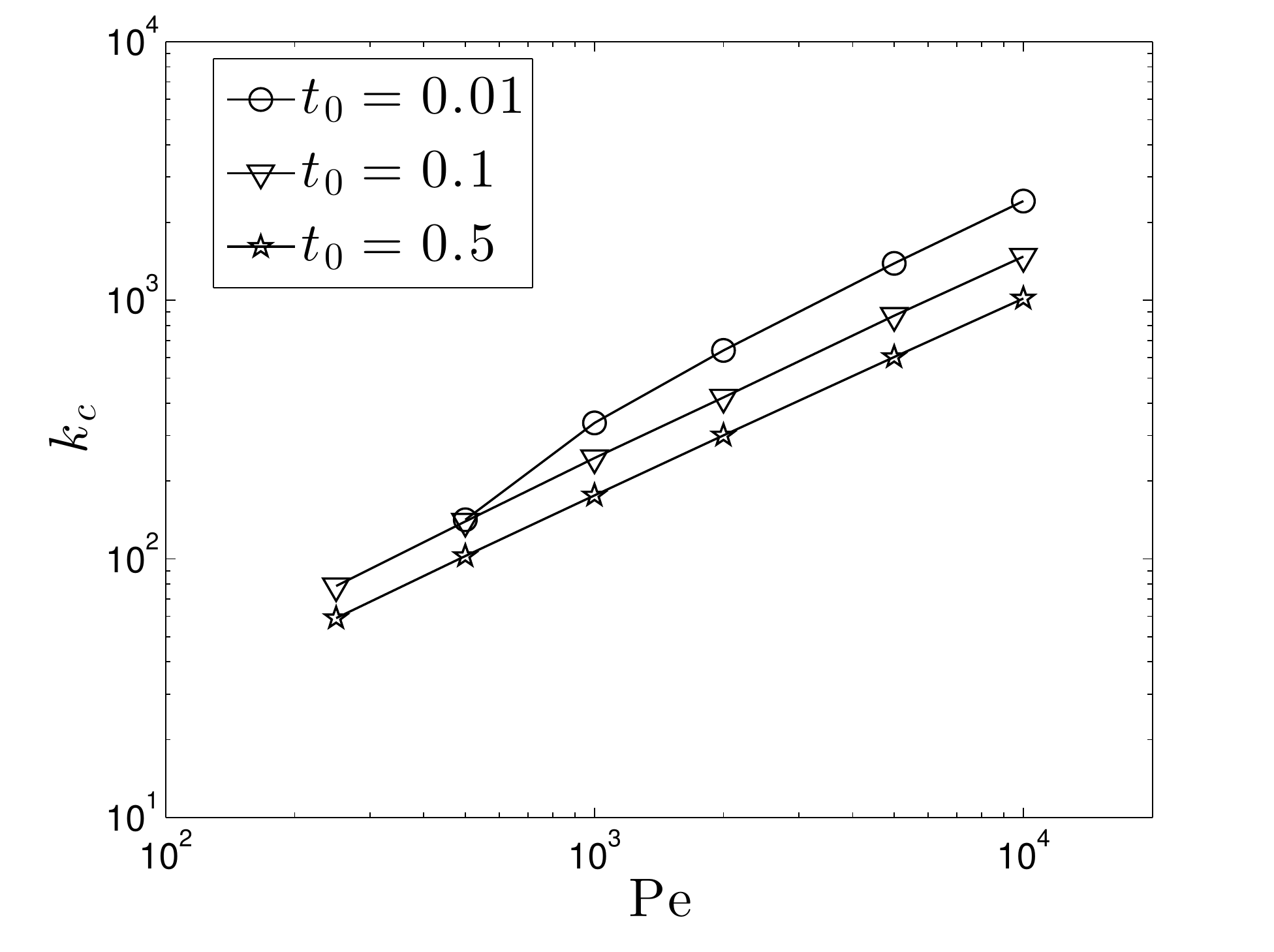}
\includegraphics[width=3.2in, keepaspectratio=true, angle=0]{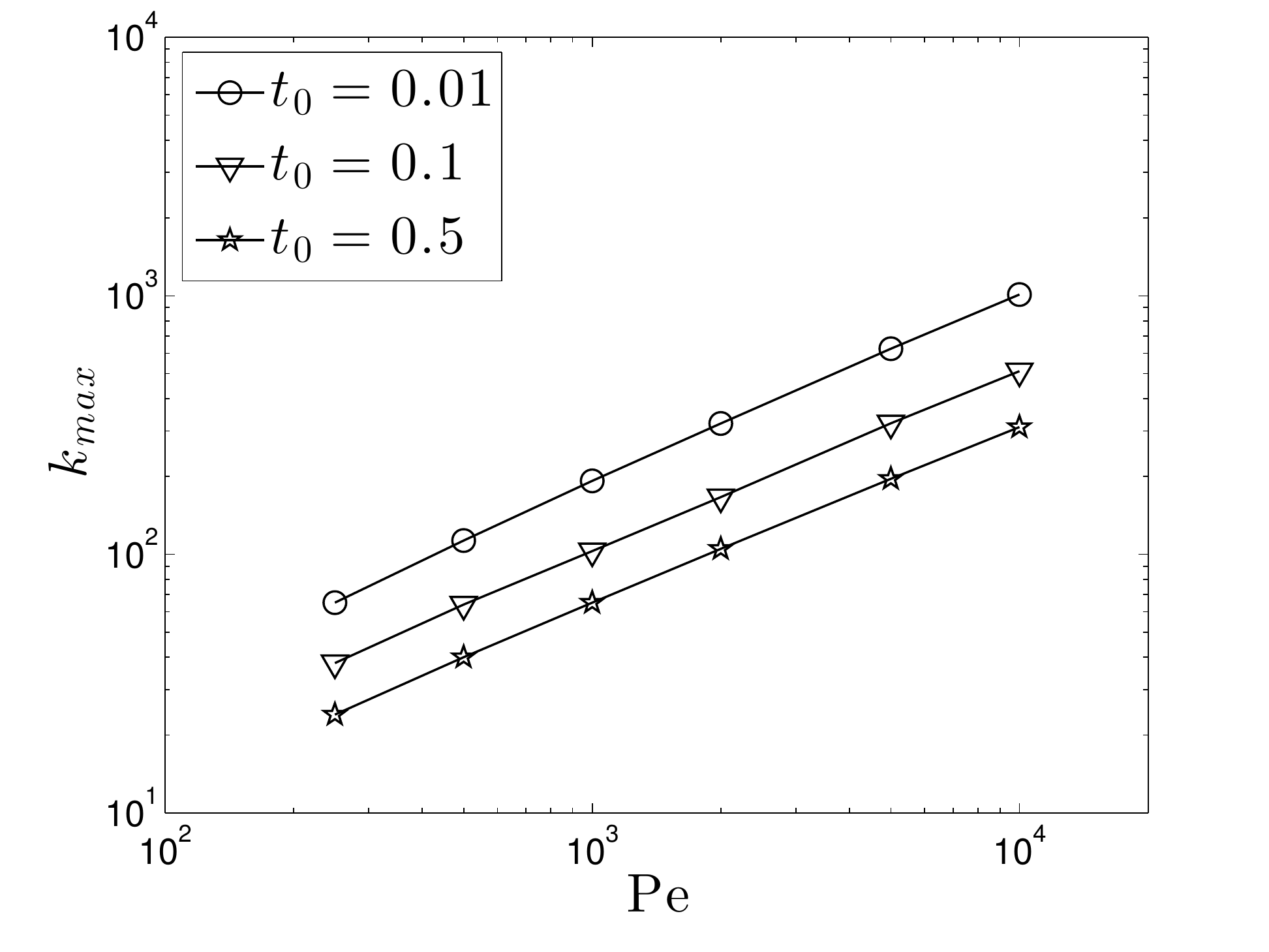}
\caption{(a) Cutoff wave number $k_c$ and (b) the most dangerous wave number $k_{max}$ as a function of Pe for $R = 3$; $k_c \sim \text{Pe}^{0.78}$ and  $k_{max} \sim \text{Pe}^{0.7}$. }\label{fig:cutoffPe}
\end{figure}

\begin{figure}[ht]
(a) \hspace{8 cm} (b) \\
\centering
\includegraphics[width=3.2in, keepaspectratio=true, angle=0]{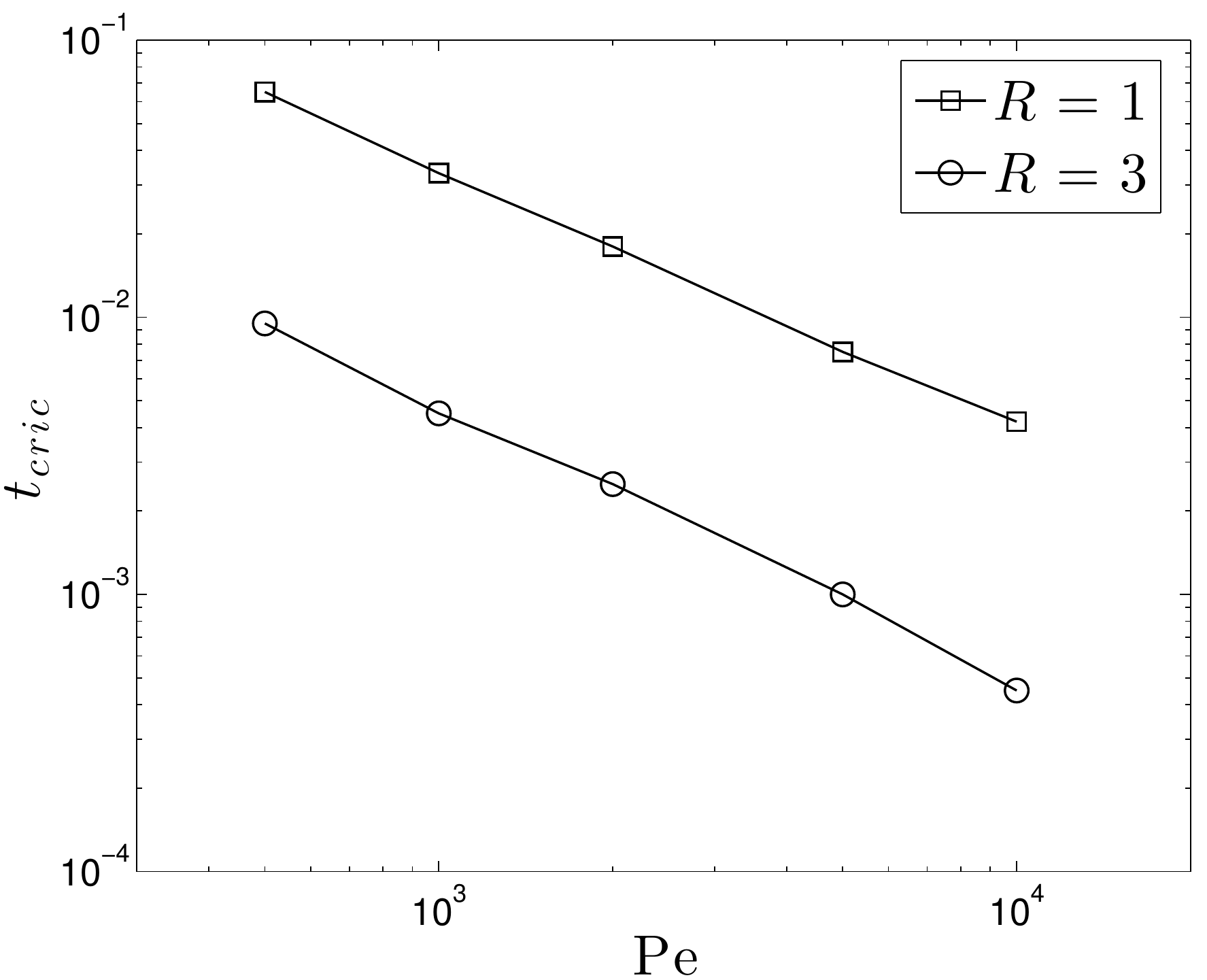}
\includegraphics[width=3.2in, keepaspectratio=true, angle=0]{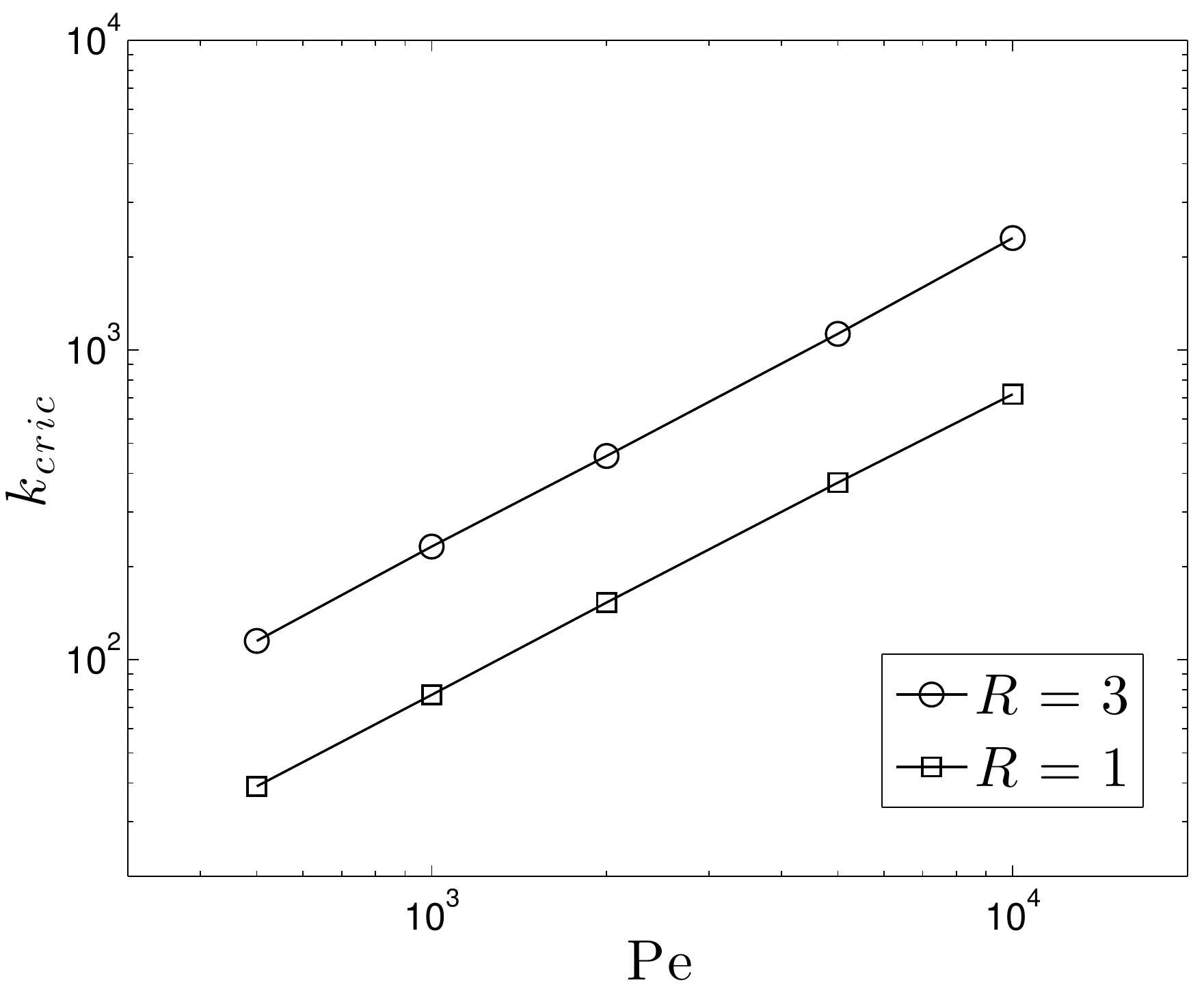}
\caption{(a) Critical time $t_{cric}$ and (b) critical wave number $k_{cric}$ as a function of Pe for different values of  $R$ with $\epsilon = 1$. It is observed that $t_{cric} \sim \text{Pe}^{-1}$ and  $k_{cric} \sim \text{Pe}$. }\label{fig:criticalPe}
\end{figure}

Figs. \ref{fig:cutofft0}(a) and \ref{fig:cutofft0}(b) represent the variations of $k_c$ and $k_{max}$, respectively, with $t_0$ for different values of Pe and $R = 3$. It has been observed that $k_c$ seems to decrease slower ($k_c \sim t_0^{-0.22}$) than that of $k_{max}$ ($k_{max} \sim t_0^{-0.3}$). As mentioned earlier, $k_c$ and $k_{max}$ increase with Pe for all values of $t_0$ and log-mobility ratio $R $; we look to investigate the rate at which they change with Pe for different values of $R$ and $t_0$. Variations of $k_c$ and $k_{max}$ with Pe for $R = 3$ and different values of $t_0$ has been plotted in Figs. \ref{fig:cutoffPe}(a) and \ref{fig:cutoffPe}(b), respectively. It is observed that $k_c$ increases more rapidly than $k_{max}$; $k_c$ seems to increase with Pe at the rate of $\text{Pe}^{0.78}$, whereas $k_{max}$ changes with Pe as $\text{Pe}^{0.7}$. The curves in logarithmic scales are straight lines with constant slope except a sudden decrease near Pe $= 10^3$ in the curve of $k_c$ for $t_0 = 0.01$ and it does not pass through Pe = $250$ (see Fig. \ref{fig:cutoffPe}(a)). However, there is no such observations for $k_{max}$. This signifies that all the perturbations at $t_0 = 0.01$ for Pe $= 250$ remains stable, which was clear from the dispersion curves (see Fig. \ref{fig:dispersion}(a)). One of the salient features of linear stability analyses is to determine the onset of instability, which is difficult to determine accurately from the direct numerical simulations and experiments. Hence, the critical time for the onset of instability $t_{cric}$ has been determined for different Pe which is found to vary as Pe$^{-1}$, and it is presented graphically in Fig. \ref{fig:criticalPe}(a). Information about the critical P\'{e}clet, Pe$_c$, for different $t_0$ will be helpful from the experimental point of view, so that Pe can be used as a control parameter with a fixed $t_0$. Such information can be easily obtained from the neutral stability curves shown in Fig. \ref{fig:neutral}, which clearly depicts that as $t_0$ increases the Pe$_c$ decreases. Fig. \ref{fig:criticalPe}(b) depicts a linear growth of the corresponding critical wave number $k_{cric}$ with Pe.

%figure6

\begin{figure}[ht]
\centering
\includegraphics[width=5in, keepaspectratio=true, angle=0]{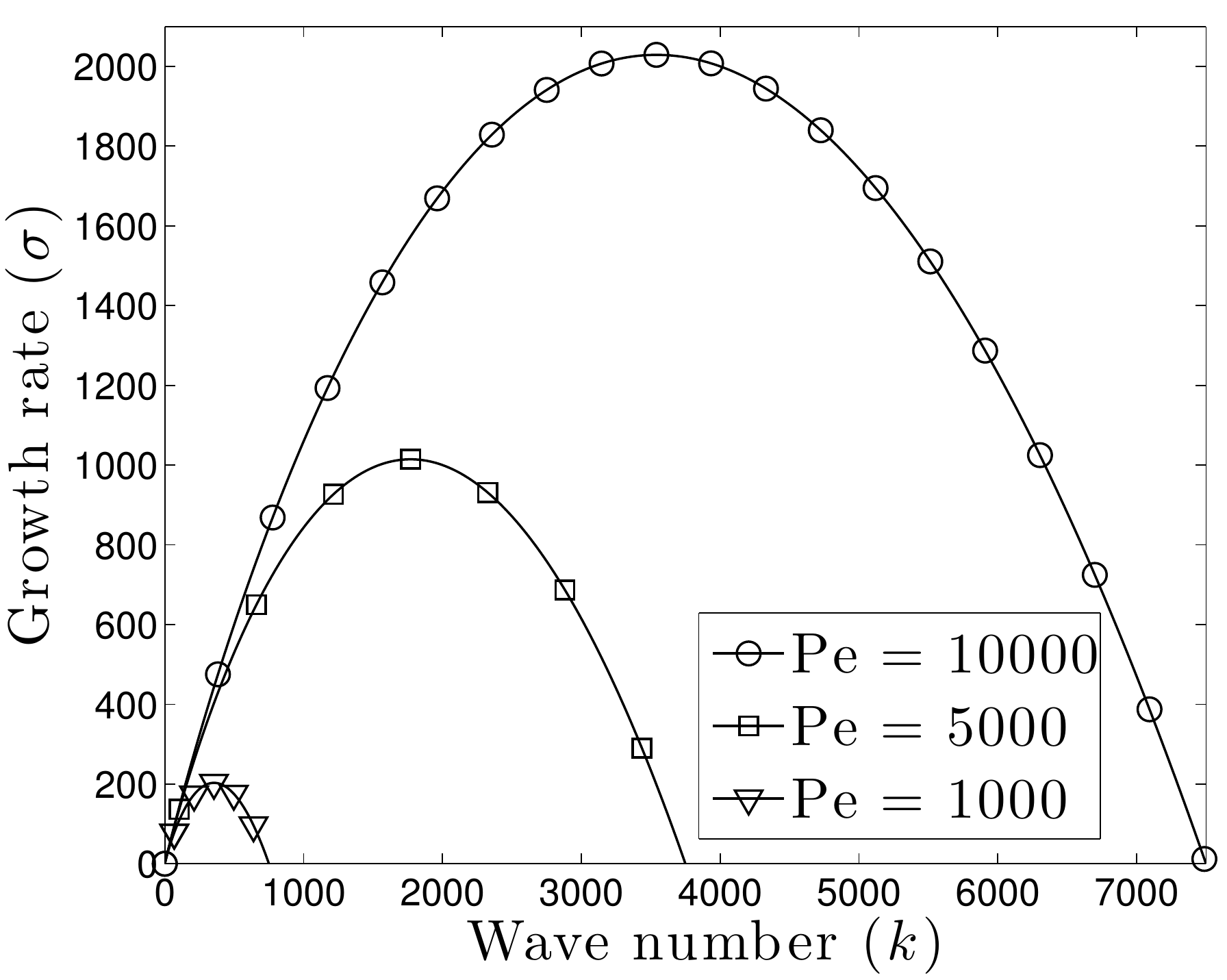}
\caption{Initial growth rate $\sigma$ of step-like concentration distribution as a function of the perturbation wave number $k$ for different Pe values (see Eq. \ref{eq:A8}). }\label{fig:ADC}
\end{figure}

In order to validate the qualitative and quantitative influence of Pe on VF instability, an analytical dispersion relation for a step-like initial concentration distribution has been computed in the $(x,y,t)$-coordinate systems and is presented in Appendix \ref{sec:ADR}. It has been found that $\sigma$ is an increasing function of Pe (see Eq. \eqref{eq:A9}), signifying a larger growth rate of a perturbation with particular wave number, as diffusion becomes slower. Fig. \ref{fig:ADC} depicts the qualitative as well as quantitative behavior of $\sigma$ with Pe, which are in good agreement with the growth rates $\sigma^*$ computed numerically in self-similar $(\xi, y, t)$-coordinate system for $t_0 > 0$ (see Fig. \ref{fig:dispersion}), as $t_0 = 0$ is a singular point in the similarity coordinate systems $(\xi,y,t)$. At time $t_0 = 0$, in the $(x,y,t)$-coordinate systems, the most unstable wave number and the corresponding growth rate are found to be, $k_{max} = (2\sqrt{5} - 4) R \;\text{Pe}/4 \approx 0.118 R\; \text{Pe}$ and $\sigma_{max} \approx 0.0225 R^2  \text{Pe}^2$, respectively. Finally the variation of the cut-off wave number with Pe has been determined by equating the growth rate to zero, which leads to $k_c = R\; \text{Pe}/4.$ Thus, it is found that both the cut-off wave number and the most dangerous wave number vary linearly with the log-mobility ratio $R$ as well as Pe at initial time $t_0 = 0$. 

%figure7

In order to determine how the variations of these two vital wave numbers with Pe changes for different frozen diffusive time $t_0$, numerical solutions are obtained for the stability equations in the $(x, y, t)$-coordinate system at $t_0 \neq 0$. It has been observed that for $R = 3$ at two different frozen diffusive times $t_0 = 0.01$ and $0.1$, $k_{max} \propto \text{Pe}^{0.35}$ and $k_c \propto \text{Pe}^{0.5}$. This signifies that as time progresses, the diffusive base state concentration looses its steepness and hence the dependence of $k_{max}$ and $k_c$ on Pe become weaker than those at initial time $t_0 = 0$. 

\subsection{Comparison of the rectilinear flow with the radial source flow}\label{subsec:CRSF}
Tan and Homsy \cite{TH3} mentioned that the principal difference between the rectilinear and the radial displacement is best described in terms of their base state velocities and the growth of the perturbations in the respective cases. In the case of the former the base state velocity is a constant, while for the radial displacement, the velocity varies spatially. This results into an algebraic growth of the perturbations in the latter case as opposed to exponential in time for the rectilinear flow. This propels us to investigate how the fluid dispersion drives various vital parameters, such as $k_t, k_{max}, k_c$, etc. for these two potentially different displacement processes in porous media. 

There lies an appreciable difference between the radial source flow and the rectilinear flow in porous media that can be readily mentioned is the temporal evolution of the critical P\'{e}clet number; in the case of the latter, Pe$_c$ varies with time unlike the case of the former where it is independent of time for a fixed value of $R$. In the rectilinear displacement, the triplet of the threshold, most dangerous and cut-off wave numbers, $(k_t, k_{max}, k_c)$, is found to vary with the time as opposed to their time independence for the radial source flow \cite{TH3}. It is mentioned earlier that $k_{max}$ and $k_c$ are found to increase with Pe in the rectilinear displacement similar to the radial flow. However, in the case of the radial source flow $k_t$ is also found to change with Pe; in fact, it decreases as Pe increases and hence allows larger wavelengths to be preferable for fingering instability. 

Recall, in Fig. \ref{fig:cutoffPe} it has been shown that $k_c \propto \text{Pe}^{0.78}$ and $k_{max} \propto \text{Pe}^{0.7}$. The rates of variation of these two vital parameters with Pe for the rectilinear displacement are in good agreement with those for radial source flow, where $k_{c} \propto \text{Pe}^{0.8}$ and $k_{max} \propto \text{Pe}^{0.71}$ \cite{TH3}. Thus it is observed that although the radial and the rectilinear displacements are potentially two different mechanisms, both the qualitative and quantitative influences of Pe on their stability are almost identical. This result arises since the growth of these two displacement processes differ in time, not with the dispersion of the systems. 

Thus linear stability analysis elucidates some interesting results about the influence of Pe number and $t_0$ on miscible viscous fingering in the rectilinear displacement in a Hele-Shaw cell, discussed in \ref{subsec:NR}. Nevertheless, LSA can only predict the growth of an infinitesimal perturbation at the diffusive fluid interface; not the interactions of the fingers and the instability patterns. Sec. \ref{sec:NS} addresses this aspect and compares the results with those of LSA.

%figure8
\begin{figure}[ht]
(a) \hspace{8 cm} (b) \\
\centering
\includegraphics[width=3.2in, keepaspectratio=true, angle=0]{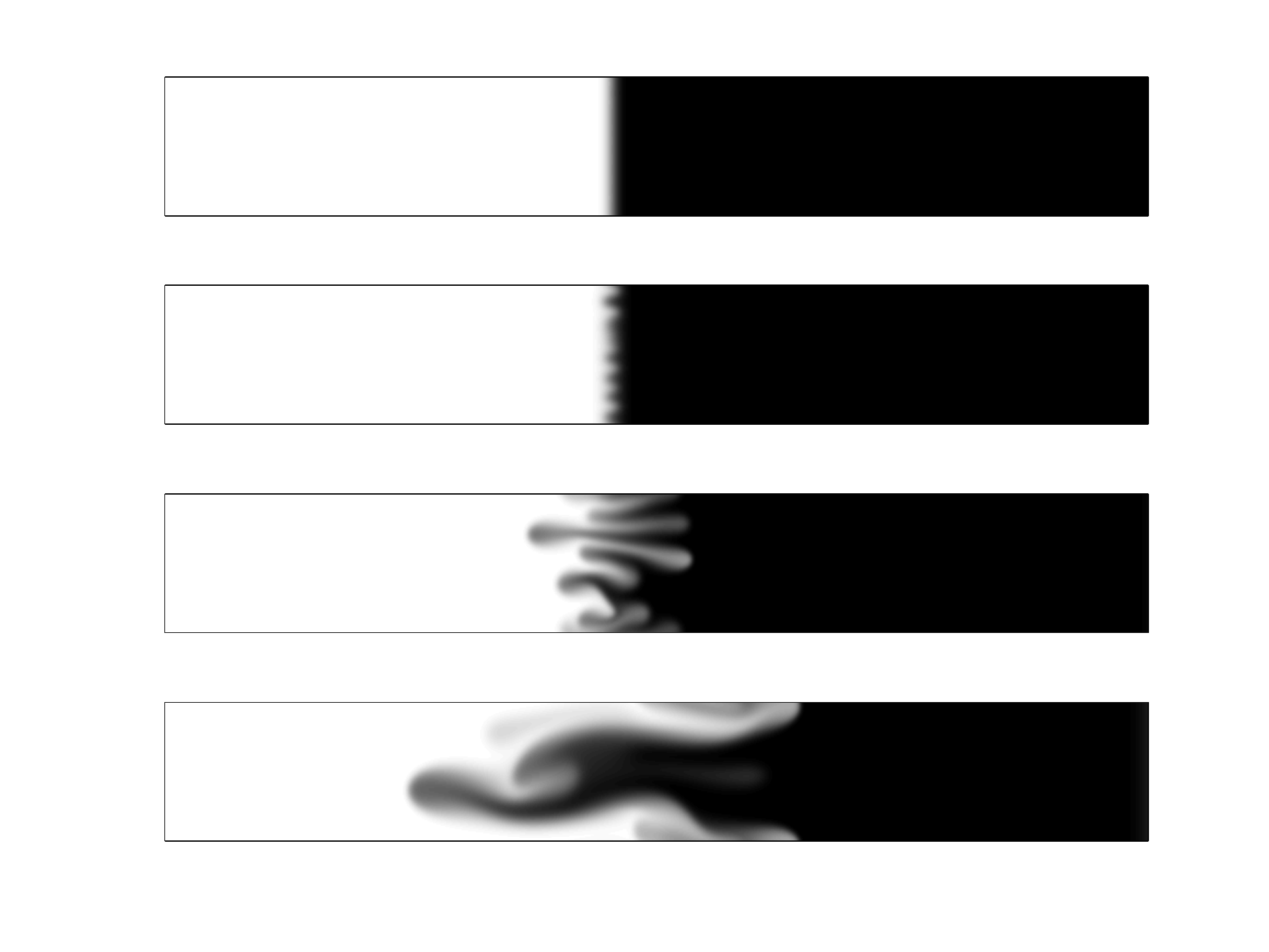}
\includegraphics[width=3.2in, keepaspectratio=true, angle=0]{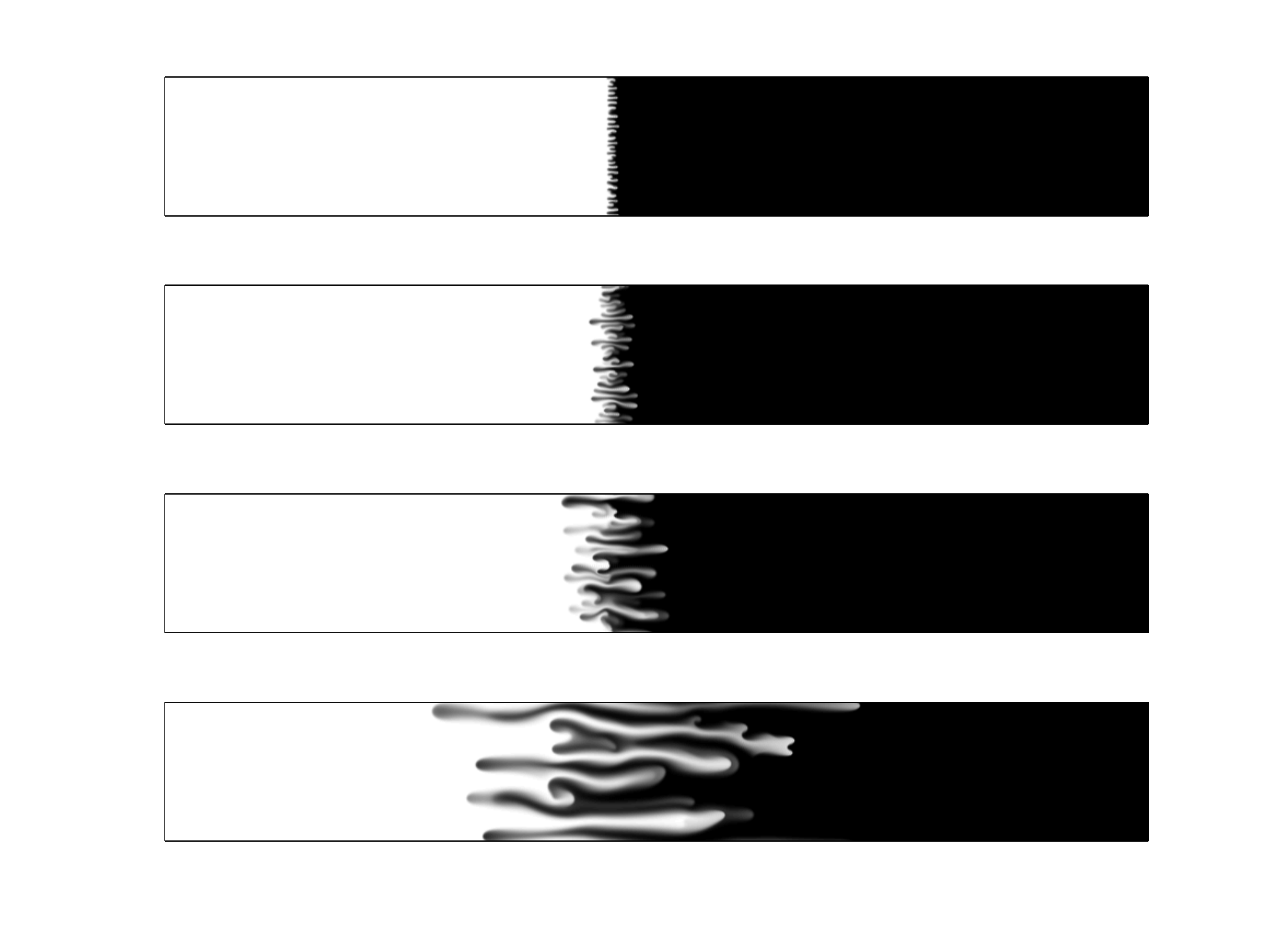}
\caption{(a) Density plots of concentration at successive times in a frame moving with the speed of the displacing fluid with $R = 1, \epsilon = 1$, (a) Pe = $1000$, from top to bottom: $t = 1, 2, 5, 10$; (b)  Pe = $5000$, from top to bottom: $t = 0.5, 1, 2, 5$. }\label{fig:density}
\end{figure}

\section{Nonlinear simulations}\label{sec:NS}
Here we look to investigate how the fluid dispersion influences the long time behavior of the nonlinear VF interactions. In order to achieve this goal, numerical simulations have been performed for the fully coupled nonlinear system of equations, Eqs. \eqref{eq:dimcon} - \eqref{eq:dimcondiff}, along with the boundary and initial conditions \eqref{eq:bc1} - \eqref{eq:initial}. Stream function-vorticity form of the dimensionless Eqs. \eqref{eq:dimcon} - \eqref{eq:dimcondiff}, 
\begin{eqnarray}
& & \nabla^2\psi = -\omega, \\
& & \frac{\partial c}{\partial t} + \frac{\partial \psi}{\partial y}\frac{\partial c}{\partial x} - \frac{\partial \psi}{\partial x}\frac{\partial c}{\partial y} = \frac{1}{\text{Pe}}\left(\frac{\partial^2c}{\partial x^2} + \epsilon\frac{\partial^2c}{\partial y^2}\right), \\
& & \omega = R\left(\frac{\partial \psi}{\partial x}\frac{\partial c}{\partial x} + \frac{\partial \psi}{\partial y}\frac{\partial c}{\partial y} + \frac{\partial c}{\partial y}\right ),
\end{eqnarray}
are solved using Fourier-spectral method following various authors (see Tan and Homsy \cite{TH2}, De Wit \textit{et al.} \cite{DBM}). The computational domain is chosen to be $[0, L_x/L_y] \times [0, 1]$ with $1024 \times 128$ spectra incorporated within a rectangle of aspect ratio $8:1$. The time integration has been performed by taking time stepping $\Delta t = 10^{-4}$. Numerical simulations are performed with $1160 \times 160$ spectra in the same computational domain and $\Delta t = 10^{-4}$. Results obtained from these two sets of simulations are in good agreement with maximum relative error of $\mathbf{O}(10^{-2})$. Due to the convergence issue of the numerical scheme we restrict ourselves to                                                                                                                                                                                                                                                                                                                                                                                                                                                                                                                                                                                                                                                                                                                                                                                                                                                                                                                                                                                                                                                                                                                                                                                                                                                                                                                                                                                                                                                                                                                                                                                                                                                                                                                                                                                                                                                                                                                                                                                                                                                                                                                                                                                                                                                                                                                                                                                                                                                                                                                                                                                                                                                                                                                                                                                                                                                                                          $R \leq 2$, Pe $\leq 5000$ and $0.2 \leq \epsilon \leq 1$, and analyze the effect of fluid dispersion on miscible VF. 

Fig. \ref{fig:density} shows density plots of the concentration field at successive dimensionless times in a reference frame moving with the speed of the displacing fluid for different values of Pe with $R = 1$ and $ \epsilon = 1$. It depicts that VF instability takes place earlier for the case of small dispersion, i.e. when Pe is higher. In particular, for Pe $= 10^3$ no visible fingers are formed at time $t =1$ (see Fig. \ref{fig:density}(a)), whereas at the same time fingers are found to form with a higher Pe $= 5000$ (see Fig. \ref{fig:density}(b)). It also shows that for smaller dispersion more number of fine scale fingers are formed, which leads to highly nonlinear interaction of the fingers, giving rise to more complex dynamical patterns. Both the qualitative and quantitative effects of Pe on VF will be discussed further in Secs. \ref{subsec:Mixing} and \ref{subsec:Onset and wave length}.

%figure9
\begin{figure}[ht]
\centering
\includegraphics[width=5in, keepaspectratio=true, angle=0]{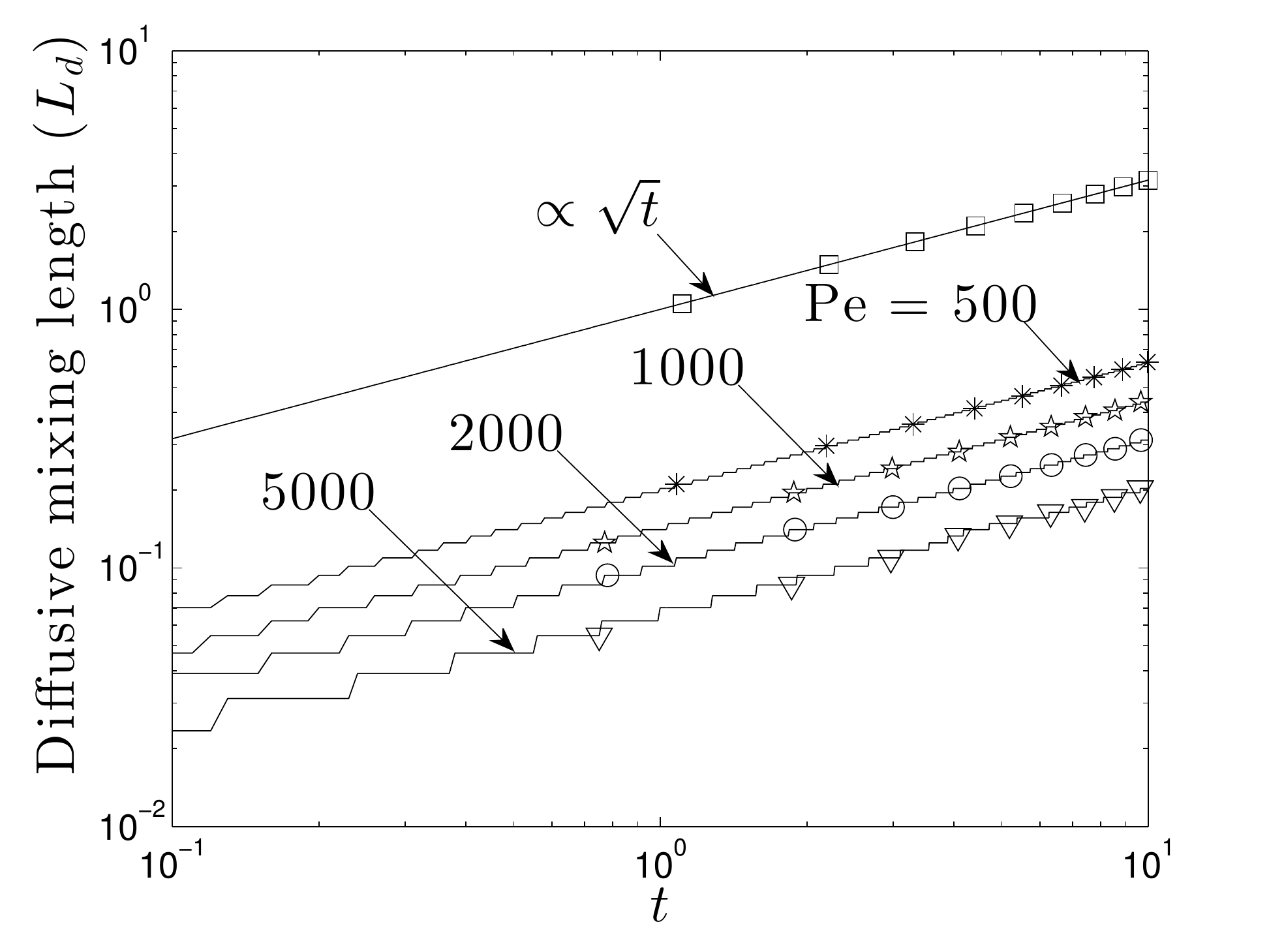}
\caption{Mixing length ($L_d \propto \sqrt{t}$) in the absence of viscous fingering instability ($R = 0$) for different values of Pe. Largest Pe (slowest diffusion) has the minimum mixing length. }\label{fig:diffmixing}
\end{figure}

\subsection{Influence of Pe on mixing lengths}\label{subsec:Mixing}
In the absence of VF ($R = 0$), mixing length $L_d$ measures the spread of the fluid region over which two fluids get mixed \cite{DBM, MMD}. Fig. \ref{fig:diffmixing} shows the evolution of $L_d$'s for different Pe values in logarithmic scales. It depicts that for all values of Pe, mixing lengths are proportional to $\sqrt{t}$ and the proportionality constant decreases with Pe. Here we determine the dependence of the proportionality constant $C_{diff}$ with Pe, such that $L_d = C_{diff}\sqrt{t}$. It has been observed that $C_{diff}$ scales as $\text{Pe}^{-1/2}$, signifying that the fluids with largest Pe have the narrowest mixing zone. Such a variation of $L_d$ with Pe is pragmatic, since we know that the diffusive mixing length varies as the square root of fluid dispersion coefficient, here $\displaystyle\frac{1}{\text{Pe}}$; therefore $L_d$ should vary as $\text{Pe}^{-1/2}$, same as predicted from the numerical simulations. Hence, all the $L_d$'s corresponding to different Pe number, after rescaling with $\text{Pe}^{-1/2}$, are found to be lying on a single straight line of slope $1/2$ in the logarithmic scales (not shown here). Henceforth, for convenient we rescale all the mixing lengths with $\text{Pe}^{-1/2}$ and perseude further discussions. 

%figure10
\begin{figure}[ht]
(a) \hspace{8 cm} (b) \\
\centering
\includegraphics[width=3.2in, keepaspectratio=true, angle=0]{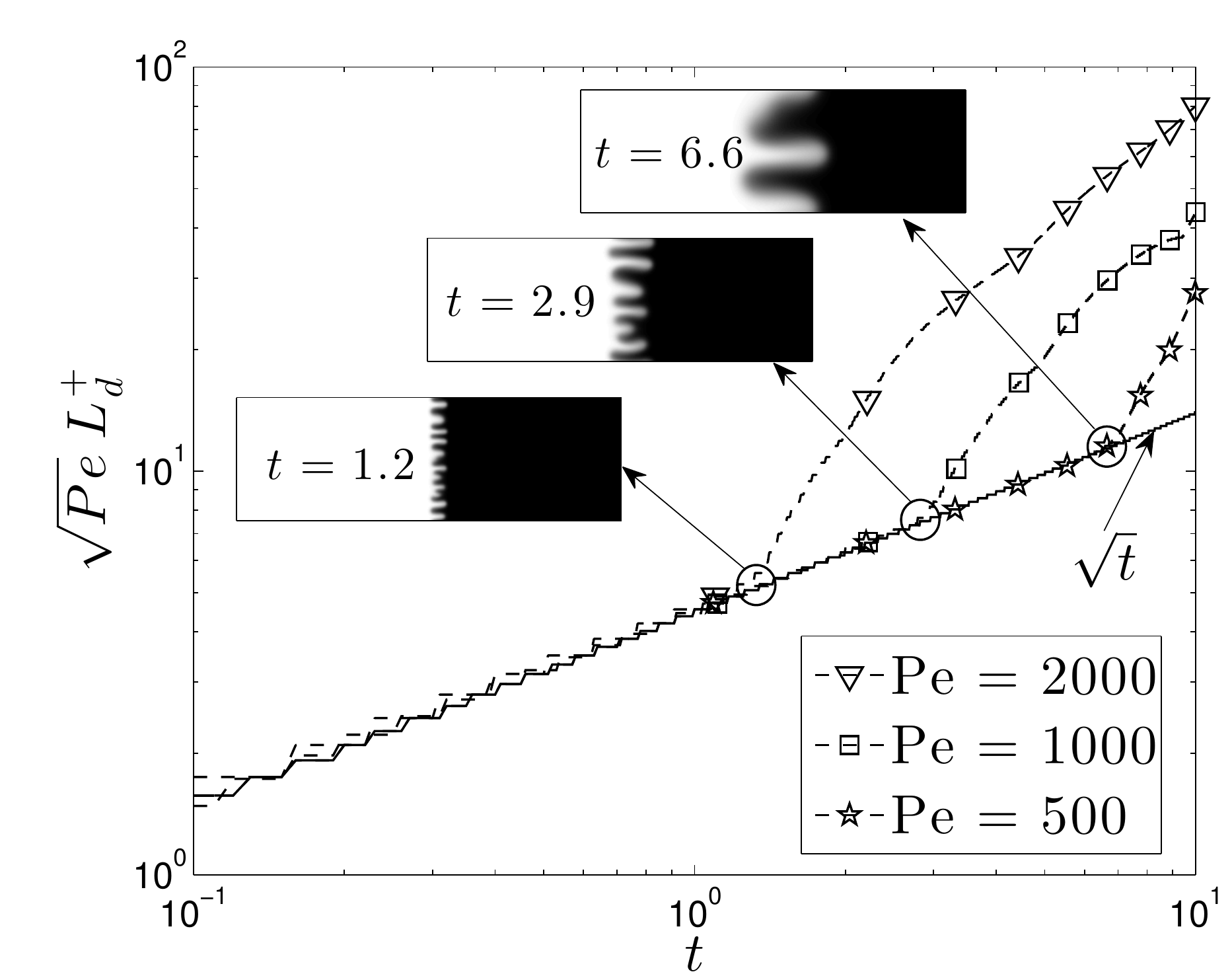}
\includegraphics[width=3.2in, keepaspectratio=true, angle=0]{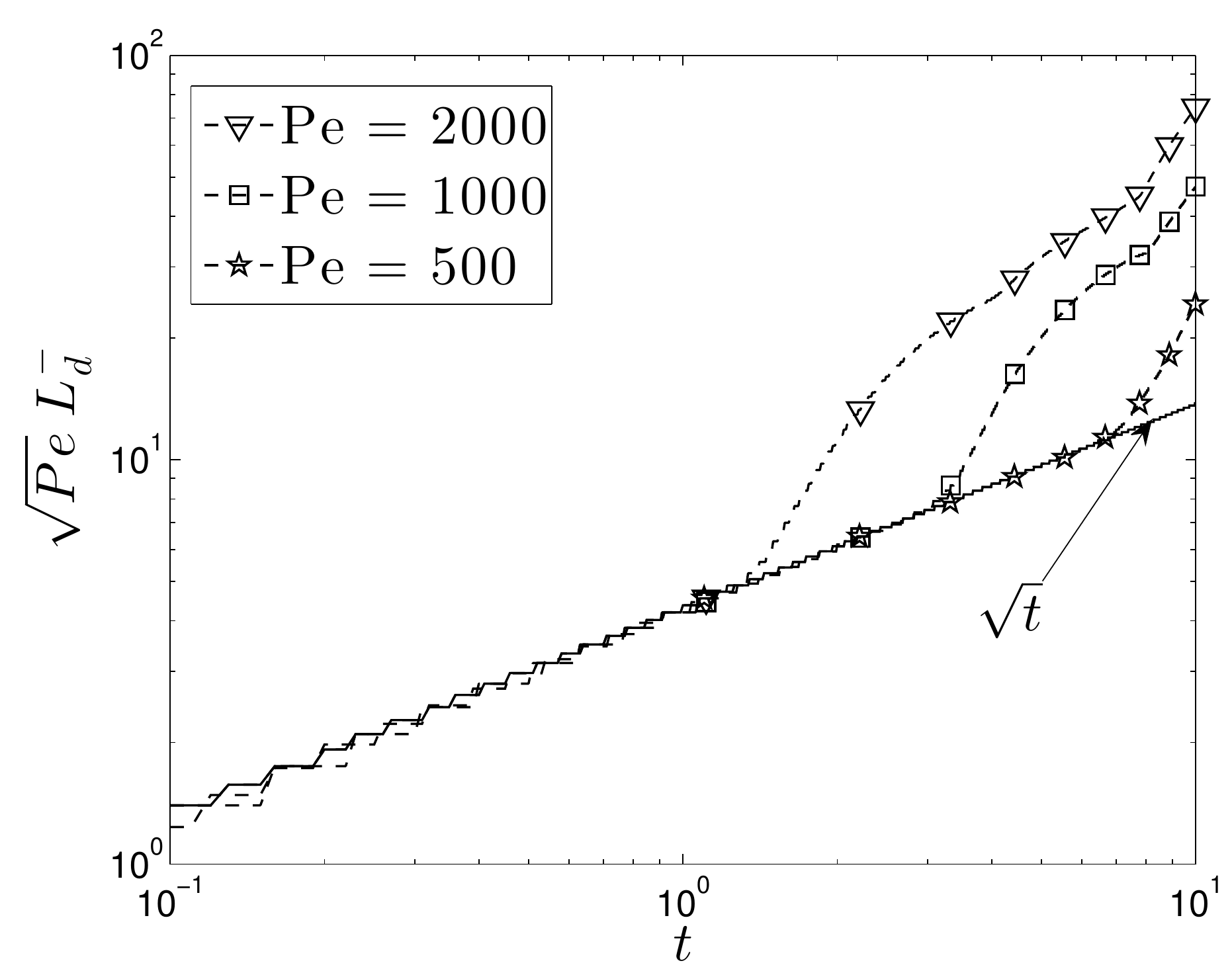}
\caption{(a) Positive mixing length ($L_d^+$) and (b) negative mixing length ($L_d^-$) for different values of Pe, scaled with $\text{Pe}^{-1/2}$, for $R = 1, \epsilon = 1$. Deviation from the purely diffusive mixing length happens earliest for the slowest diffusive system, i.e. for the case of highest Pe value. The corresponding time and the distribution of the concentration field are depicted in the insets of (a), showing the fingers. }\label{fig:VFmixing}
\end{figure}

In the presence of the VF instability two different mixing lengths are discussed in the literature;\cite{MH} positive and negative mixing lengths, $L_d^+$ and $L_d^-$, respectively, which typically measure the lengths of the fingers in the downstream and upstream directions. Next, we analyze the influence of Pe on the rescaled positive and negative mixing lengths, $\sqrt{\text{Pe}}~L_d^+$ and $\sqrt{\text{Pe}}~L_d^-$, which are presented in Figs. \ref{fig:VFmixing}(a) and \ref{fig:VFmixing}(b), respectively. They depict that at an early time both $L_d^+$ and $L_d^-$ follow diffusive scaling and depending upon Pe, after a particular time they propagate with a different power law higher than $1/2$. Both $L_d^+$ and $L_d^-$ are seen to propagate as $t^2$ immediate after the onset of fingering for all values of Pe. Nevertheless, $L_d^+$ sustains this power law for a relatively longer period than $L_d^-$, as the latter one is formed in the upstream direction against the flow. This change in the propagation rate of the mixing lengths are due to the fingering effects and the instant, when the mixing lengths deviate from diffusive scale to fingering scale, seems to be the onset of nonlinear fingers. However, it has been observed that the fingers start to appear even before that particular instant, but either of the mixing lengths fails to capture it properly as these fingers are of very small amplitude and are confined within the diffusive layers. Nevertheless, the interfacial length,\cite{MMD} $I(t) = \displaystyle\int_0^1\int_{0}^{L_x}\left[\left(\frac{\partial c}{\partial x}\right)^2\ + \left(\frac{\partial c}{\partial y}\right)^2\right]^{1/2}\text{d}x\text{d}y$, captures the onset of nonlinear fingers, $t_{on}$, more appropriately. This is defined as the time when $I(t)$ increases $10\%$ from its constant value in the diffusive regime, which equals the dimensionless width of the computational domain. As expected, systems with slower dispersion have earlier onset, since in this case the stabilization due to dispersion is the less and hence fingers form earlier.

%figure11
\begin{figure}[ht]
(a) \hspace{8 cm} (b) \\
\centering
\includegraphics[width=3.2in, keepaspectratio=true, angle=0]{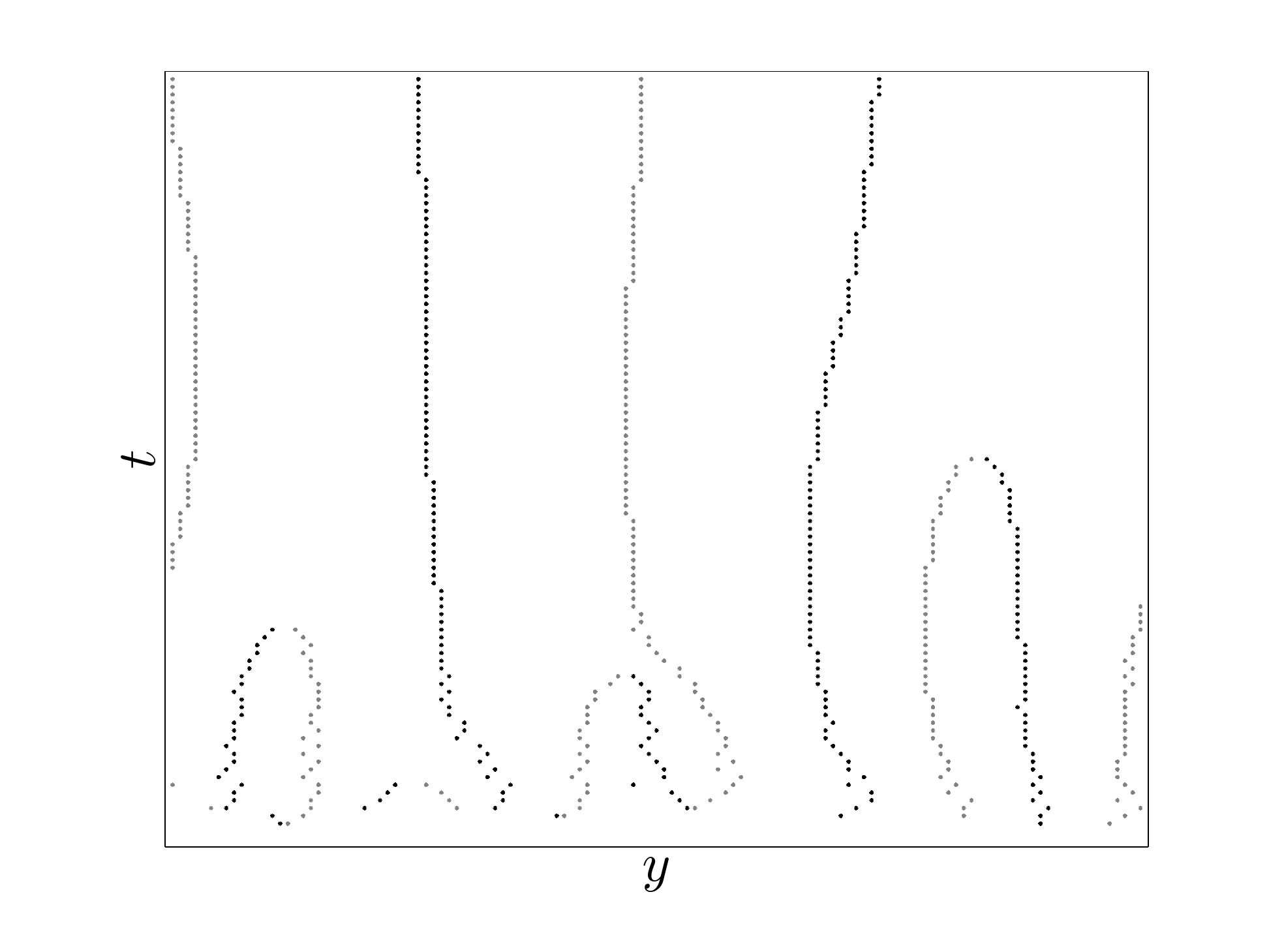}
\includegraphics[width=3.2in, keepaspectratio=true, angle=0]{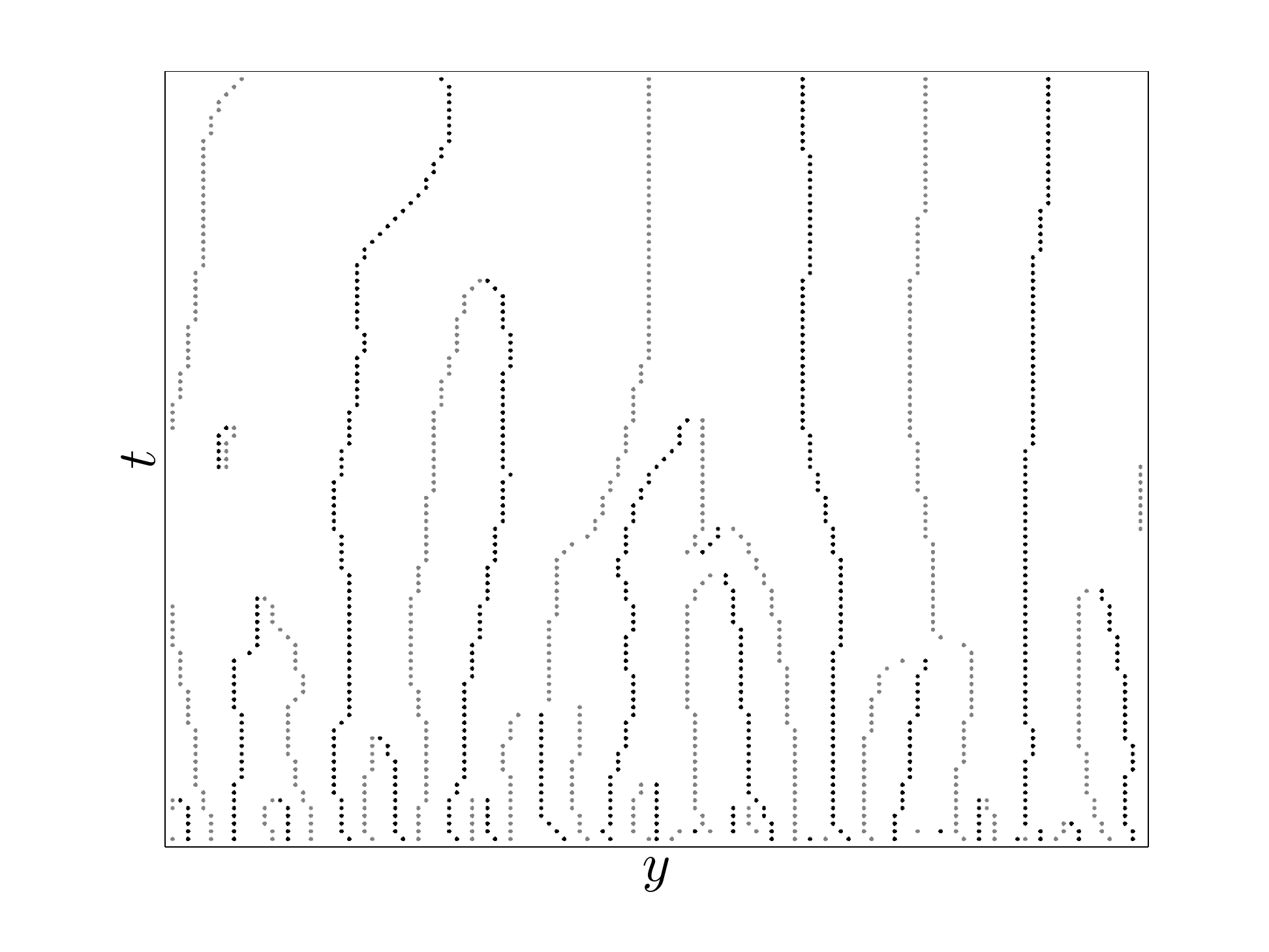}
\caption{Space-time maps of the locations of the local maxima (black) and minima (gray) of the longitudinal averaged concentration profile $\bar{c}(y,t)$ as a function of time for the case of $R = 1, \epsilon = 1$ with (a) Pe = $500$ and (b) Pe = $2000$. The dimensionless time $t$ varies from $0$ to $10$ in the vertical direction. }\label{fig:spacetime}
\end{figure} 

\subsection{Influence of Pe on the onset of fingering and wavelength selection}\label{subsec:Onset and wave length}
Here the focus of this study is to investigate the influence of Pe on the onset of fingering instability and wavelengths of the unstable perturbations. A good measure for the visualization of the onset of fingering instability, splitting and merging of fingers and wave number of the unstable fingers is the space-time maps of the positions of the local maxima and minima of the longitudinally averaged concentration profile, $\bar{c}(y,t) = \displaystyle\frac{1}{L_x}\int_0^{L_x}c(x,y,t)\text{d}x$. Fig. \ref{fig:spacetime} shows the spatio-temporal evolution of the locations of local maxima and minima of $\bar{c}(y,t)$ for $R = 1, \epsilon = 1$ and Pe $= 500, 2000$. It depicts that for Pe $= 500$ (see Fig. \ref{fig:spacetime}(a)) very few number of local maxima and minima are observed after leaving a void space at an early time. After almost half of the run-time of the present simulations only two pairs of local maxima and minima are observed with very little deviations of their respective positions. On the other hand, for Pe $= 2000$ (see Fig. \ref{fig:spacetime}(b)) the local maxima and minima are densely crowded for small values of $t$ without any void at early time. It further depicts that more number of local maxima and minima remain present even after sufficiently long time. These signify that in the case of large diffusion coefficient the onset of fingering takes place at a later time compared to the situation with smaller diffusion and the number of fingers is significantly less in the case of the former than the case of latter. 

%figure12
\begin{figure}[ht]
\centering
\includegraphics[width=5in, keepaspectratio=true, angle=0]{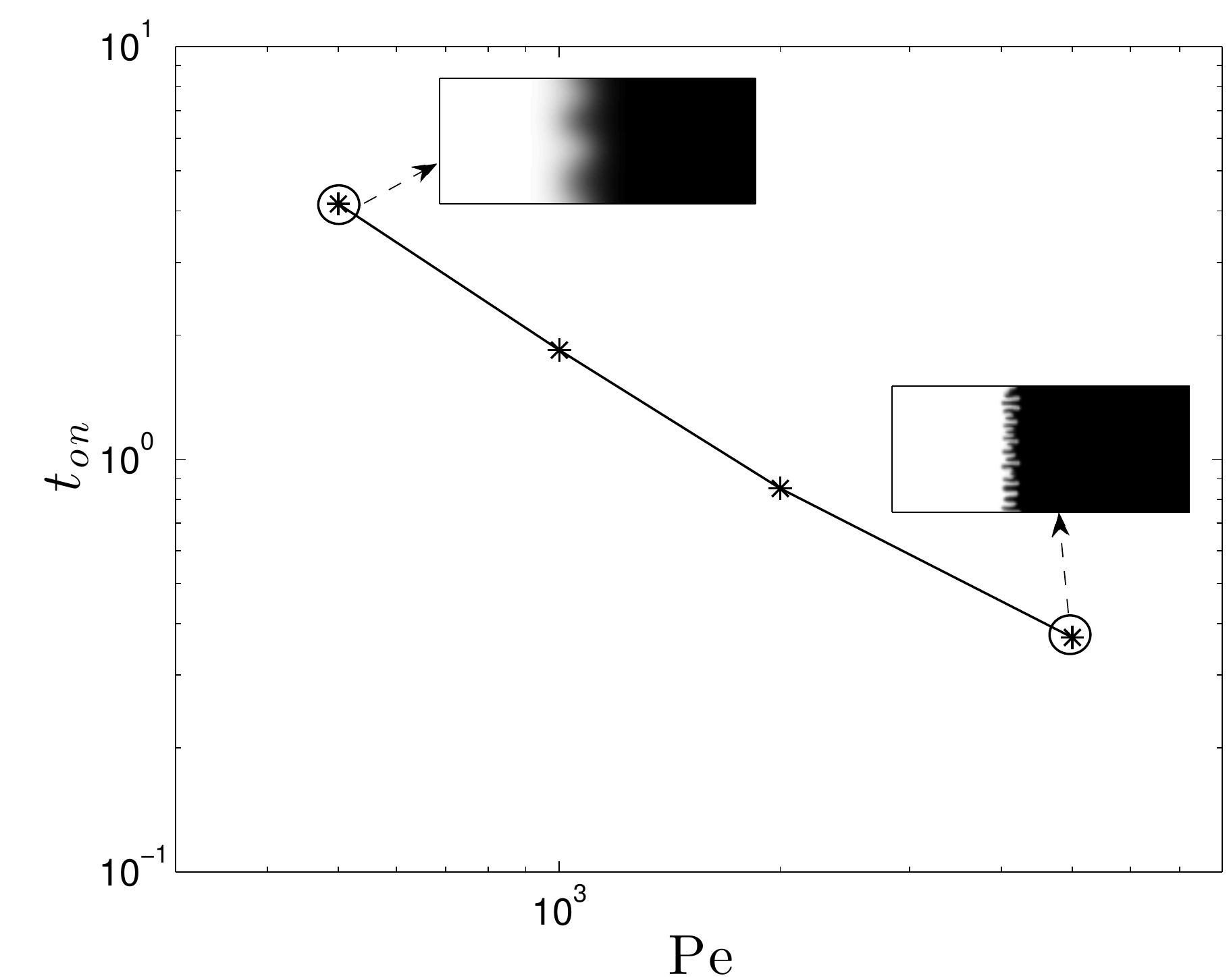}
\caption{Onset time of nonlinear fingers $t_{on}$ as a function of Pe for $R = 1, \epsilon = 1$; $t_{on} \sim \text{Pe}^{-1.1}$. }\label{fig:tonPe}
\end{figure}

Numerical simulations are performed for five different initial noise corresponding to each Pe for a fixed $R$, and then an arithmetic mean of the obtained results has been taken to get the relation between $t_{on}$ and Pe. The dependence of $t_{on}$ on Pe has been represented in Fig. \ref{fig:tonPe} that depicts that $t_{on}$ varies as $\text{Pe}^{-1.1}$ for $R = 1$. Simulations are performed with different $R$ and it turns out that such a quantitative dependence of the onset time, $t_{on}$, on Pe remains almost unaltered with different values of $R$ (not shown here). Influence of Pe can also be explained from a different point of view. From the definition of Pe $= \displaystyle\frac{UL_y}{D_x}$, it is clear that Pe is directly proportional to the displacing velocity $U$. Hence, for a fixed axial dispersion $D_x$, increase in Pe corresponds to an increase in $U$ (within the limits of Darcy velocity) and vice-verse. Taking this into consideration we conclude that increasing Pe will lead to an early onset of fingering, as it is observed in the present study. However, the dimensionless formulation with diffusive characteristic scales \cite{TH2} shows the same onset time, $t_{on}$, for different Pe, but more number of fingers which is quit reasonable as the width of the computational domain increases with Pe. Therefore, such a dimensionless model does not give an insight about the influence of fluid dispersion (i.e. Pe) on the onset of instability, which has been thoroughly investigated through the present dimensionless model. 

Irregular deflections in the space-time map of the local maxima and minima correspond to the splitting and merging of fingers. It is observed that in the case of small values of Pe merging is dominant than splitting (see Fig. \ref{fig:spacetime}(a)). However, for large values of Pe occasionally merging is followed by splitting of fingers and they further merge very quickly among themselves or with other neighboring fingers (see Fig. \ref{fig:spacetime}(b)). It further manifests that for a small Pe value, frozen fingers of constant wavelengths are formed early; whereas for higher Pe, strong and complex interaction of the fingers persists for sufficiently large time that retard to forming frozen fingers of constant wavelengths. 

%figure13
\begin{figure}[ht]
(a) \hspace{8 cm} (b) \\
\centering
\includegraphics[width=3.2in, keepaspectratio=true, angle=0]{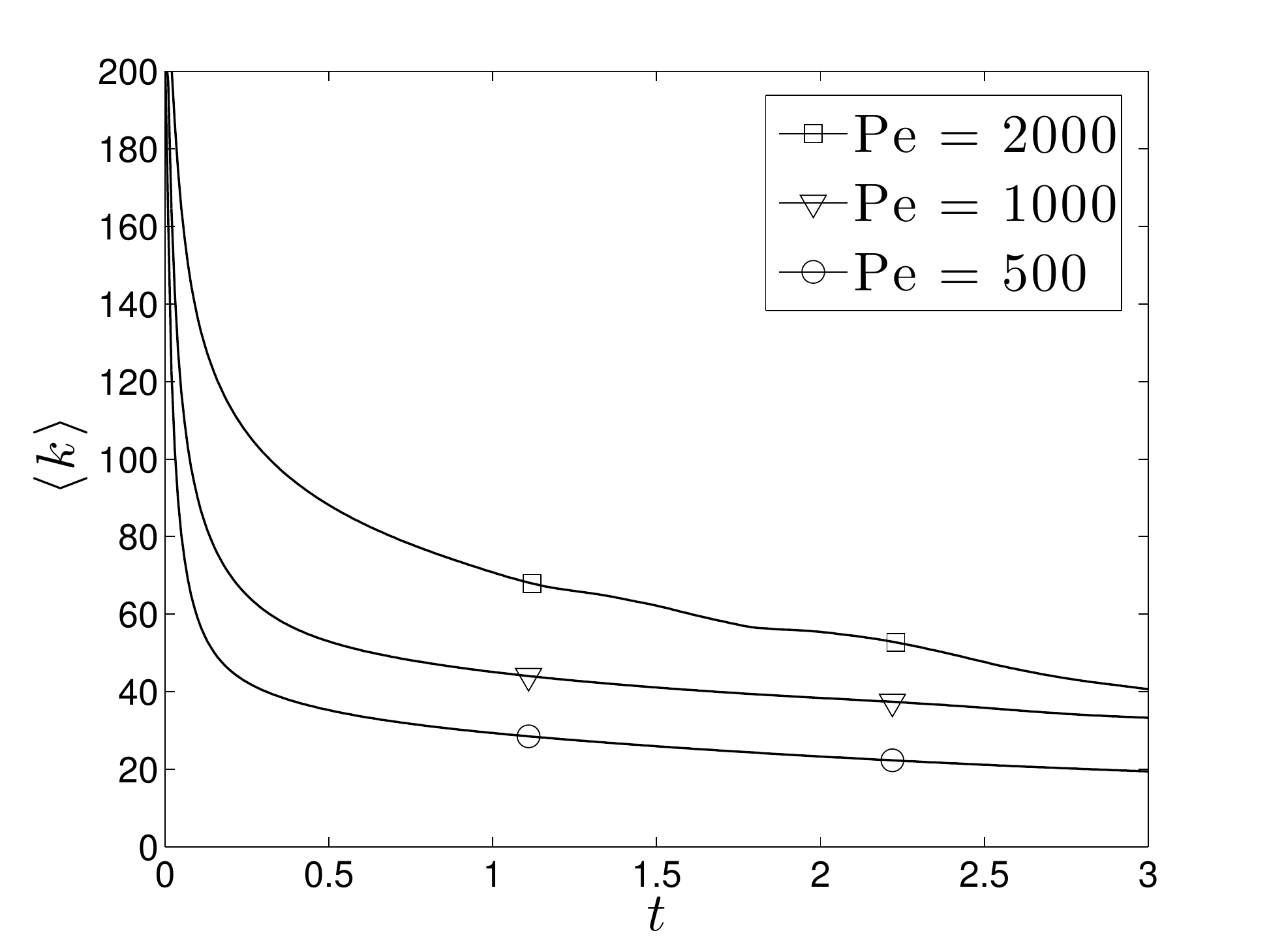}
\includegraphics[width=3.2in, keepaspectratio=true, angle=0]{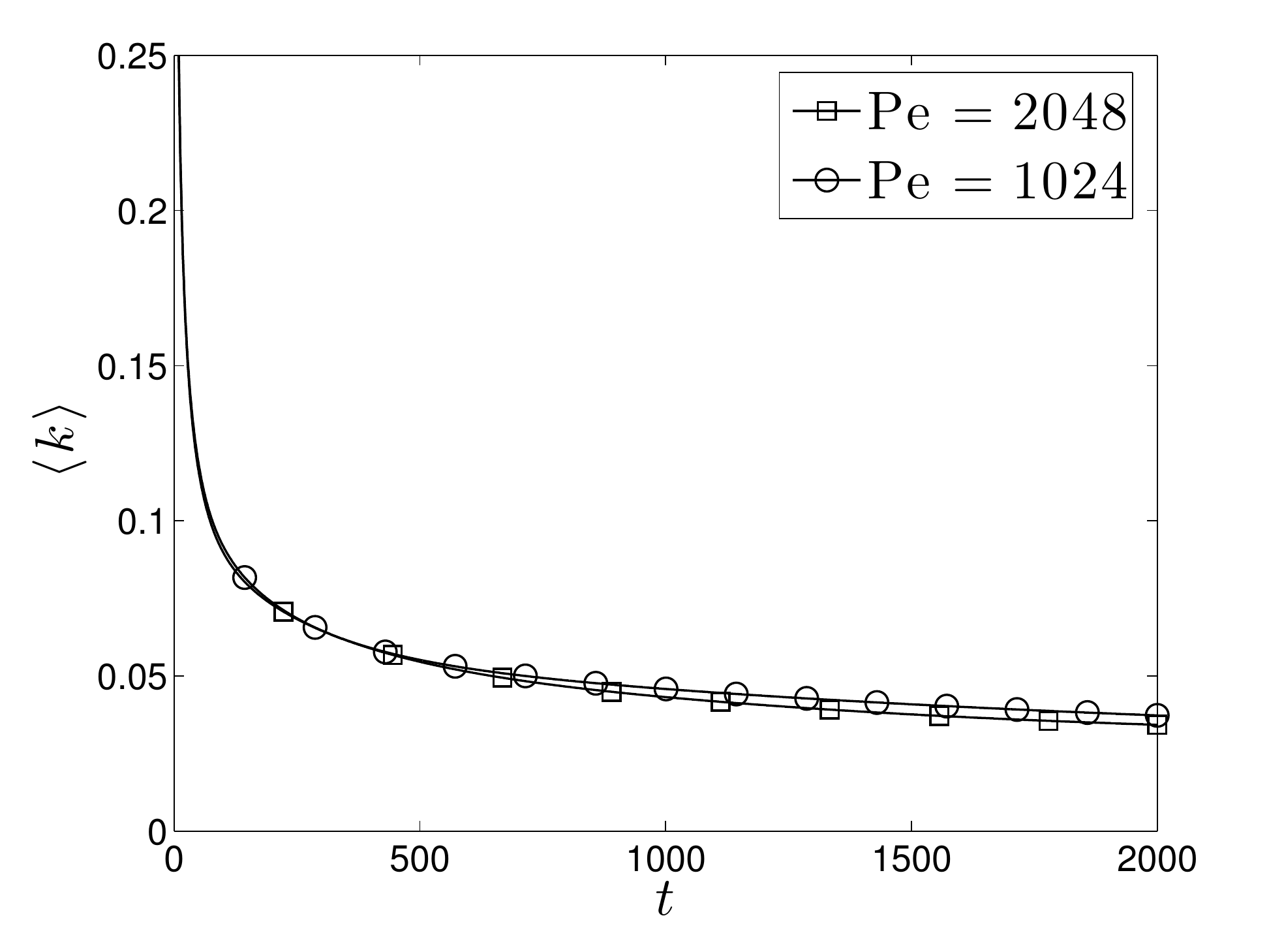}
\caption{Average wave number $\langle k \rangle$ as a function of time for different Pe values with $R = 1, \epsilon = 1$: Nondimensional formulation with (a) convective characteristic length scale, (b) diffusive characteristic length\cite{TH2}. In the latter case $\langle k \rangle$ appear to be almost identical for both the Pe values.}\label{fig:averagewavenumber}
\end{figure}

Wavelengths of the unstable disturbances can also be measured in terms of the averaged wave number of the longitudinally averaged concentration profile $\bar{c}(y,t)$. The temporal evolution of the average wave numbers $\langle k \rangle$ of the unstable modes for different values of Pe have been presented in Fig. \ref{fig:averagewavenumber} for $R = 1, \epsilon = 1$. Here, $\langle k \rangle$ is defined as $\langle k \rangle = \sum_ik_iP(k_i)/\sum_iP(k_i)$, where $k_i$ are the Fourier modes of the Fourier transform $\hat{c}(k_i,t)$ of $\bar{c}(y,t)$ and $P(k_i) = \lvert\hat{c}(k_i)\rvert$ are the amplitude in the Fourier space. It clearly depicts that at an early time $\langle k \rangle$'s differ significantly for different fluid dispersion (i.e. for different values of Pe). It also depicts that the wave number and hence the wavelength becomes constant earlier for smaller Pe values, which also confirms the results depicted in Fig. \ref{fig:spacetime}. However, this is not the case for the dimensionless models derived using diffusive characteristic length scale (see Tan and Homsy \cite{TH2}). The average wave number $\langle k \rangle$ turns out to be same for all the Pe values in these formulations. In particular, Fig. \ref{fig:averagewavenumber}(b) shows $\langle k \rangle$ corresponding to the dimensionless formulation of Pramanik and Mishra \cite{PM1} for Pe $= 1024$ and $2048$ with $R = 3, \epsilon = 1$, in the absence of the Korteweg stresses (i.e. $\delta = 0$). It is observed that $\langle k \rangle$ appears to be almost identical for both the Pe values; although the number of fingers increases with Pe, which widen the computational domain as mentioned earlier. 

%figure14
\begin{figure}[ht]
(a) \hspace{8 cm} (b) \\
\centering
\includegraphics[width=3.2in, keepaspectratio=true, angle=0]{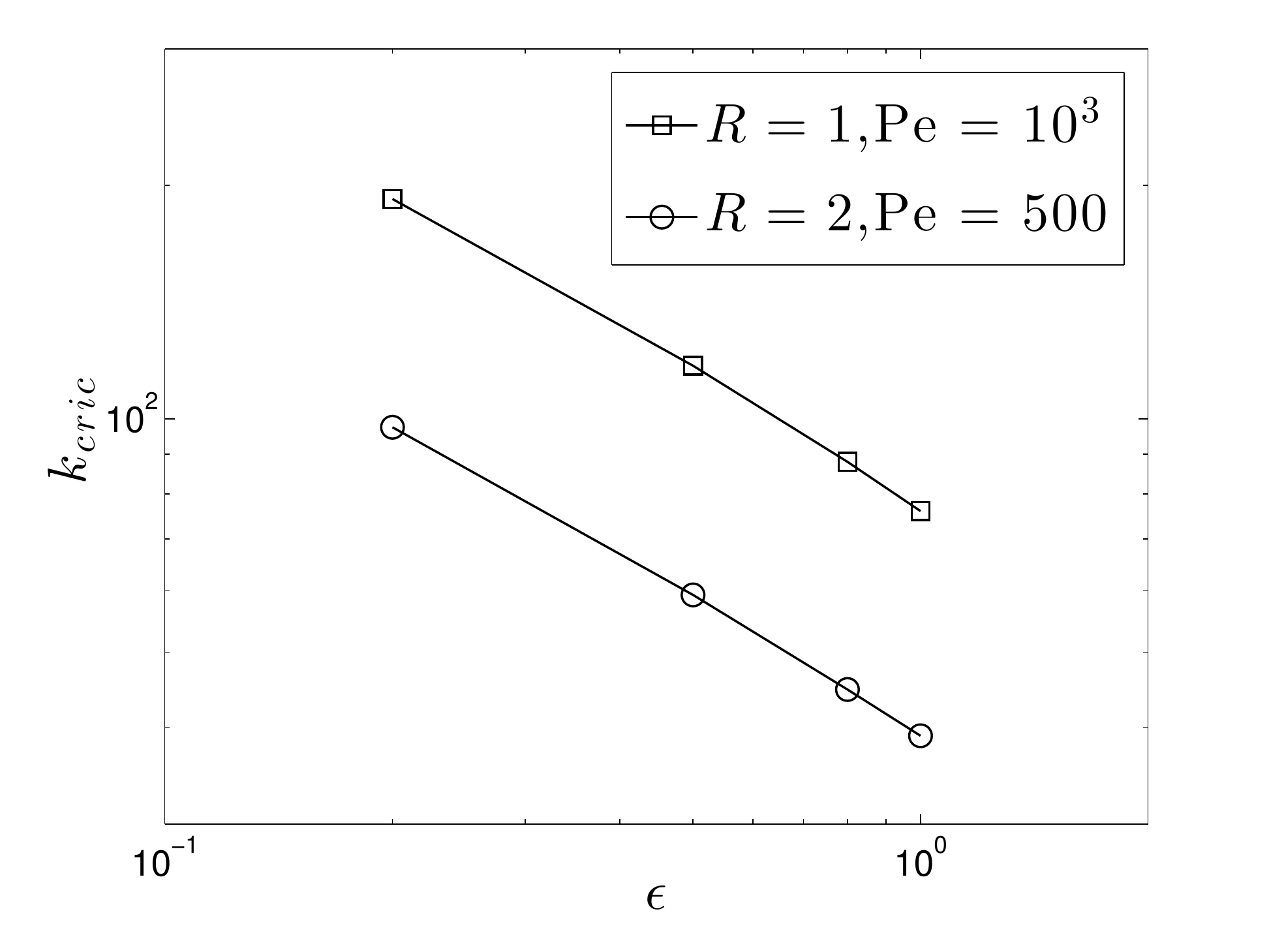}
\includegraphics[width=3.2in, keepaspectratio=true, angle=0]{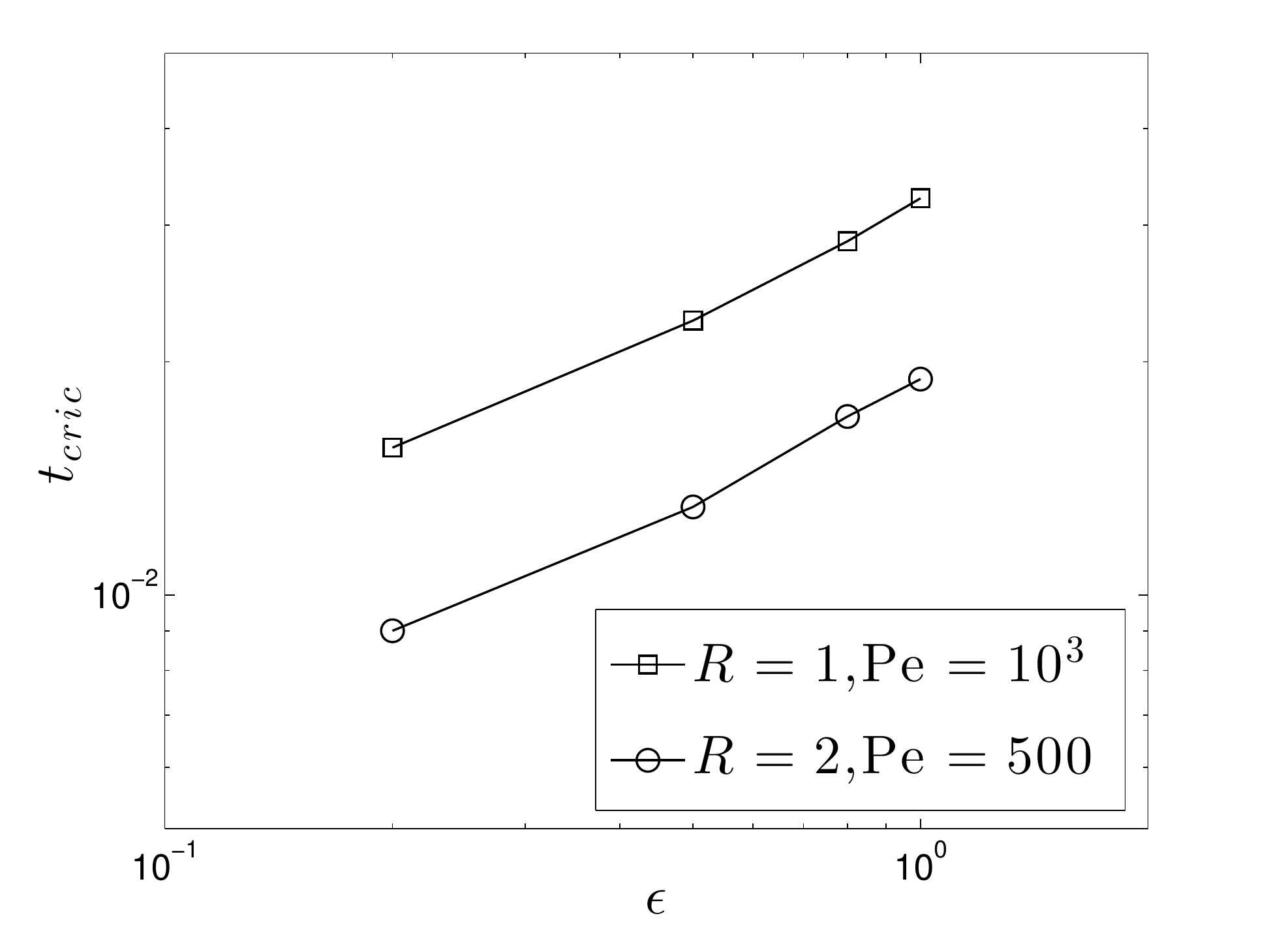}
\caption{Effect of dispersion anisotropy on (a) Critical wave number $k_{cric}$ and (b) critical time $t_{cric}$; $k_{cric}$ decreases at slightly faster rate ($k_{cric} \sim \epsilon^{-0.55}$) than the rate of increase for $t_{cric}$ ($t_{cric} \sim \epsilon^{0.5}$). }\label{fig:criticaleps} 
\end{figure}

\subsection{Effect of anisotropic dispersion}\label{subsec:AID}
The dimensionless transverse dispersion coefficient is $\displaystyle\frac{\epsilon}{\text{Pe}}$, hence by investigating the influence of $\epsilon$ and keeping Pe fixed we can trace how the anisotrpic dispersion influences VF instability both in the linear and nonlinear regimes. Influence of anisotropic dispersion in the linear regime has been analyzed in terms of two critical parameters, $k_{cric}$ and $t_{cric}$. Figs. \ref{fig:criticaleps}(a) and \ref{fig:criticaleps}(b), respectively depict the dependence of $k_{cric}$ and $t_{cric}$ on $\epsilon$ for different values of $R$ and Pe. It is observed that $k_{cric}$ decreases at a slightly faster rate ($k_{cric} \sim \epsilon^{-0.55}$) than the rate of increase of $t_{cric}$ ($t_{cric} \sim \sqrt{\epsilon}$) as $\epsilon$ increases towards the isotropic value. These variations are independent of the log-mobility ratio and the P\'{e}clet number. The eigenvalue problem (Eqs. \eqref{eq:EIG1} - \eqref{eq:EIG2}) are further solved for two extreme cases of $\epsilon \ll 1$ and $\epsilon \gg 1$. It has been observed that for $\epsilon \leq \mathbf{O}(10^{-2}) ~ k_{max} \propto \epsilon^{-0.3}$, whereas for $\epsilon \geq \mathbf{O}(10^{1}) ~ k_{max} \propto \epsilon^{-0.9}$, for fixed values of Pe and $R$ at $t_0 = 0.1$. These are in sustainable accordance with those obtained by Tan and Homsy \cite{TH1} from asymptotic analyses for $\epsilon \to 0$ and $\epsilon \to \infty$, respectively at $t_0 = 0$; in the former limiting case $k_{max}$ varies as $\epsilon^{-1/3}$ and in the latter $k_{max} \propto \epsilon^{-1}$. 

%figure15
\begin{figure}[ht]
\centering
\includegraphics[width=5in, keepaspectratio=true, angle=0]{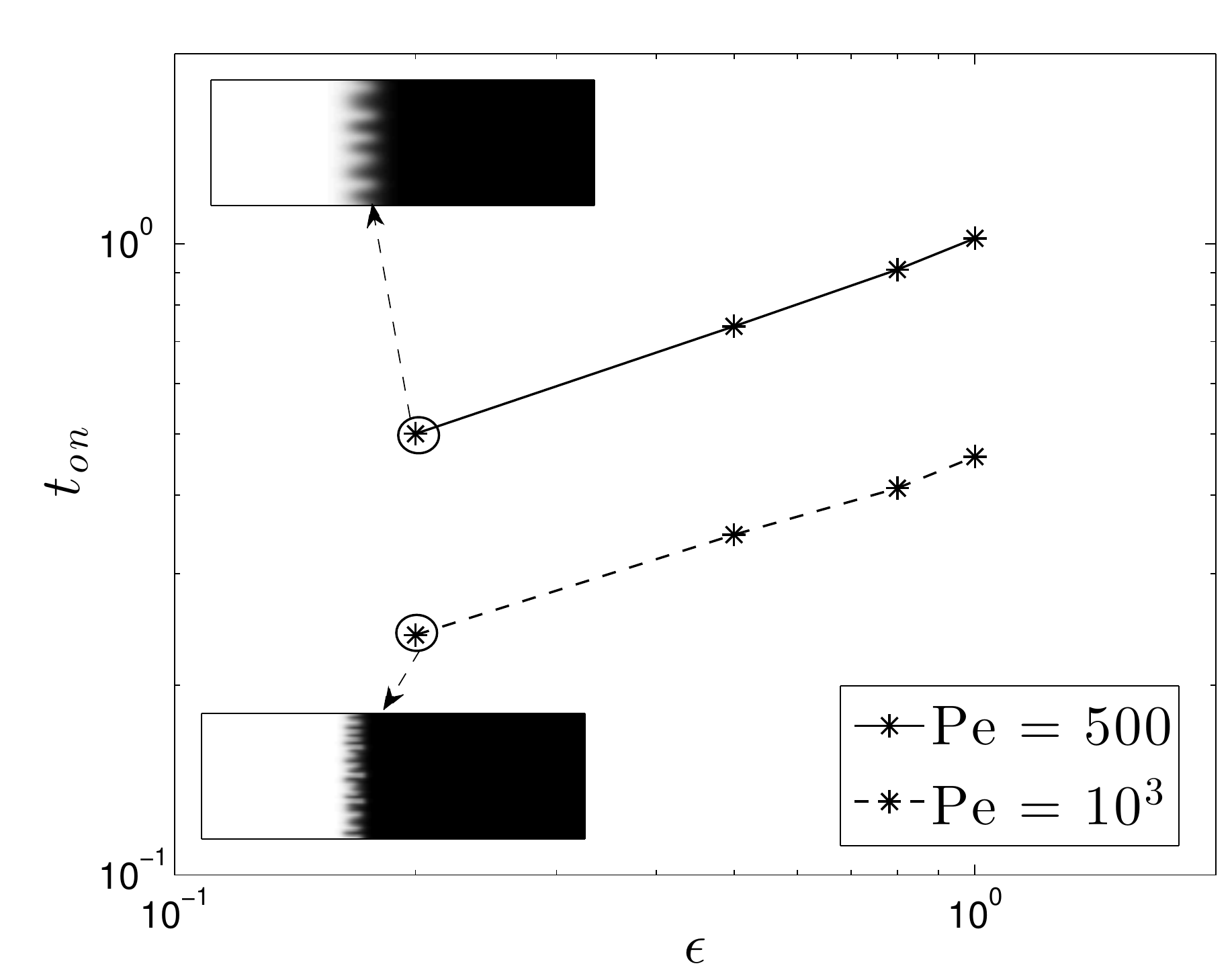}
\caption{Onset time of nonlinear fingers $t_{on}$ as a function of $\epsilon$ with $R = 2$. For both the values of Pe $t_{on} \sim \epsilon^{0.44}$. }\label{fig:toneps}
\end{figure}

Further investigation has been performed to analyze the dependence of $t_{on}$ on the transverse dispersion coefficient $\epsilon$ for a fixed Pe and $R$. Same as earlier, an arithmetic mean of the numerical simulations corresponding to the five different initial noise have been used to determine the dependence of $t_{on}$ on $\epsilon$. In Fig. \ref{fig:toneps}, onset time $t_{on}$ has been plotted against $\epsilon$ for $R = 2$ and Pe $= 500, 1000$, which clearly describes that $t_{on}$  increases with $\epsilon$ at a rate of $\epsilon^{0.44}$. This physically signifies the fact that small transverse dispersion is not only responsible for fine scale fingers, but also the onset of instability happens to be early when the transverse dispersion coefficient, $\epsilon$, is small. Thus, we observe that the influence of Pe is stronger on the onset of instability in the nonlinear regime than that in the linear regime, whereas, $\epsilon$ has a stronger impact on $t_{cric}$ than $t_{on}$. However, in general the axial dispersion coefficient has a stronger influence than the transverse diffusivity on the onset of instability both in the linear and nonlinear regimes. 

\section{Influence of P\'{e}clet number with gradient stress}\label{sec:IPeKS}
In Sec. \ref{sec:LSA} and \ref{sec:NS} it has been shown that slower dispersion rate (i.e. larger Pe) keeps the viscosity contrast of the two fluids very sharp, hence gives rise to stronger instability. Nevertheless, it has been proved both theoretically \cite{K, PM1, Polymer_Monomer_2005} and experimentally \cite{MFMG_2007} that steep concentration gradient is responsible for the Korteweg stress, which acts against the growth of the perturbations. Therefore, a sharp concentration gradient can give rise to two opposite actions, stabilization and destabilization, in miscible VF instability. Remaining of the paper will discuss the interplay between the Korteweg stresses and fluid dispersion and their subsequent effect on VF. In the presence of the Korteweg stress fluid velocity is governed by the Darcy-Korteweg equation, \cite{PM1}
\begin{equation}
\nabla P = -\mu (\vec{u} + \hat{i}) + \delta\nabla\cdot(\nabla c \otimes \nabla c).
\end{equation}
The dimensionless parameter $\delta$, defined as $\delta = \displaystyle\frac{\hat{\delta}c_2^2\kappa}{\mu_1UL_y^3}$, is called the Korteweg stress constant. Linearization of the equations and Fourier mode decomposition of the perturbation quantities, as discussed in Sec. \ref{sec:LSA}, leads to the following eigenvalue problem with Pe and $\delta$ as the flow parameters, 
\begin{eqnarray}
\label{eq:EIG1K}
\left(\frac{d^2}{d\xi^2} + R\frac{dc_0}{d\xi}\frac{d}{d\xi} - k^2t_0\right)\phi(\xi) & = & Rk^2t_0\psi(\xi) + \nonumber \\
& & ~~~ \left[\frac{k^2\delta}{\mu_0\sqrt{t_0}}\left\{\frac{d^3c_0}{d\xi^3} - \frac{dc_0}{d\xi}\left(\frac{d^2}{d\xi^2} - k^2t_0\right)\right\}\right]\psi(\xi), \\
\label{eq:EIG2K}
\left(\sigma^*(k,t_0) - \frac{1}{\text{Pe}}\left(\frac{1}{t_0}\frac{d^2}{d\xi^2} - \epsilon k^2\right)\right. & - & \left.\frac{\xi}{2t_0}\frac{d}{d\xi}\right)\psi(\xi) = -\frac{1}{\sqrt{t_0}}\frac{dc_0}{d\xi}\phi(\xi).
\end{eqnarray}
The resultant eigenvalue problem, Eqs. \eqref{eq:EIG1K} - \eqref{eq:EIG2K}, is solved numerically and the results obtained are discussed in Sec \ref{subsec:NRK}. 

\subsection{Numerical results}\label{subsec:NRK}
The objective of this part of the paper is to analyze the influence of fluid dispersion on VF in the presence of the Korteweg stress. As mentioned earlier, slow fluid dispersion helps in longer stay of the steep concentration gradient, which acts in favor of the instability and simultaneously induces the Korteweg stress that acts as a stabilizing factor in this stability problem. Hence, interplay between these two opposite influences on the fingering instability is anticipated. Such interplay has been investigated for different Pe values and the dimensionless Korteweg stress parameter $\delta$ derived from the corresponding dimensional values measured for different miscible systems by Pojman and coworkers \cite{Pojman_Langmuir_2006, Zoltowski_Langmuir_2007, MFMG_2007}. Zoltowski {\it et al.} \cite{Zoltowski_Langmuir_2007} have estimated the Korteweg stress parameter to be on the order of $10^{-11}$ N to $10^{-8}$ N for a miscible Dodecyl Acrylate-Poly(Dodecyl Acrylate) system using spinning drop tensiometry in room temperature. In another set of experiments in microgravity Pojman {\it et al.} \cite{MFMG_2007} have measured this stress parameter as $10^{-12}$ N for the honey-water system. The dimensionless value of the Korteweg stress parameter depends on the characteristic length, velocity, viscosity and the permeability of the medium. Thus, depending upon these parameters dimensionless value can take values from a wide range of interval. Typically, in Hele-Shaw cell experiments \cite{MRSMCD} or liquid chromatography \cite{Fernandez} the uniform displacement speed ranges over $10^{-5}$ ms$^{-1}$ to $10^{-2}$ ms$^{-1}$. Permeability of porous rock ranges between 1 to 100 mD, i.e. $10^{-15}$ to $10^{-13}$ m$^2$. Viscosity of water is approximately 1 cP = $10^{-3}$ N-s/m$^2$. The characteristic width of the Hele-Shaw cell is $10^{-1}$ m and the gap width between the parallel plates is $\mathbf{O}(10^{-4})$ m., which corresponds to a permeability $b^2/12$ equivalent to $10^{-7}$ m$^2$. Hence,  corresponding to the above parameter ranges the dimensionless value of Korteweg stress parameter becomes $ \sim O(10^{-16}$) to $O(10^{-4})$. The sign of the Korteweg stress parameter has been determined to be negative both theoretically \cite{Hu_ZAMP_1992} and experimentally \cite{Sugii}. Since, the maximum relative error of the pseudo-spectral method used in this paper is of order $10^{-2}$ (as mentioned in Sec. \ref{sec:NS}), the lower limit of the dimensionless parameter $\delta$ is too small to perform numerical simulations. Moreover, due to the convergence of the numerical technique, discussion of the Kortewge stress effect for a wide range of the dimensionless parameter $\delta$ is beyond the scope of this paper. Thus we have taken two different values from this parameter range in our investigation, $\delta = -10^{-5}$ and $-10^{-6}$, which are of the same order used by Chen {\it et al.} \cite{CWM} 

%figure16
\begin{figure}[ht]
\centering
\includegraphics[width=5in, keepaspectratio=true, angle=0]{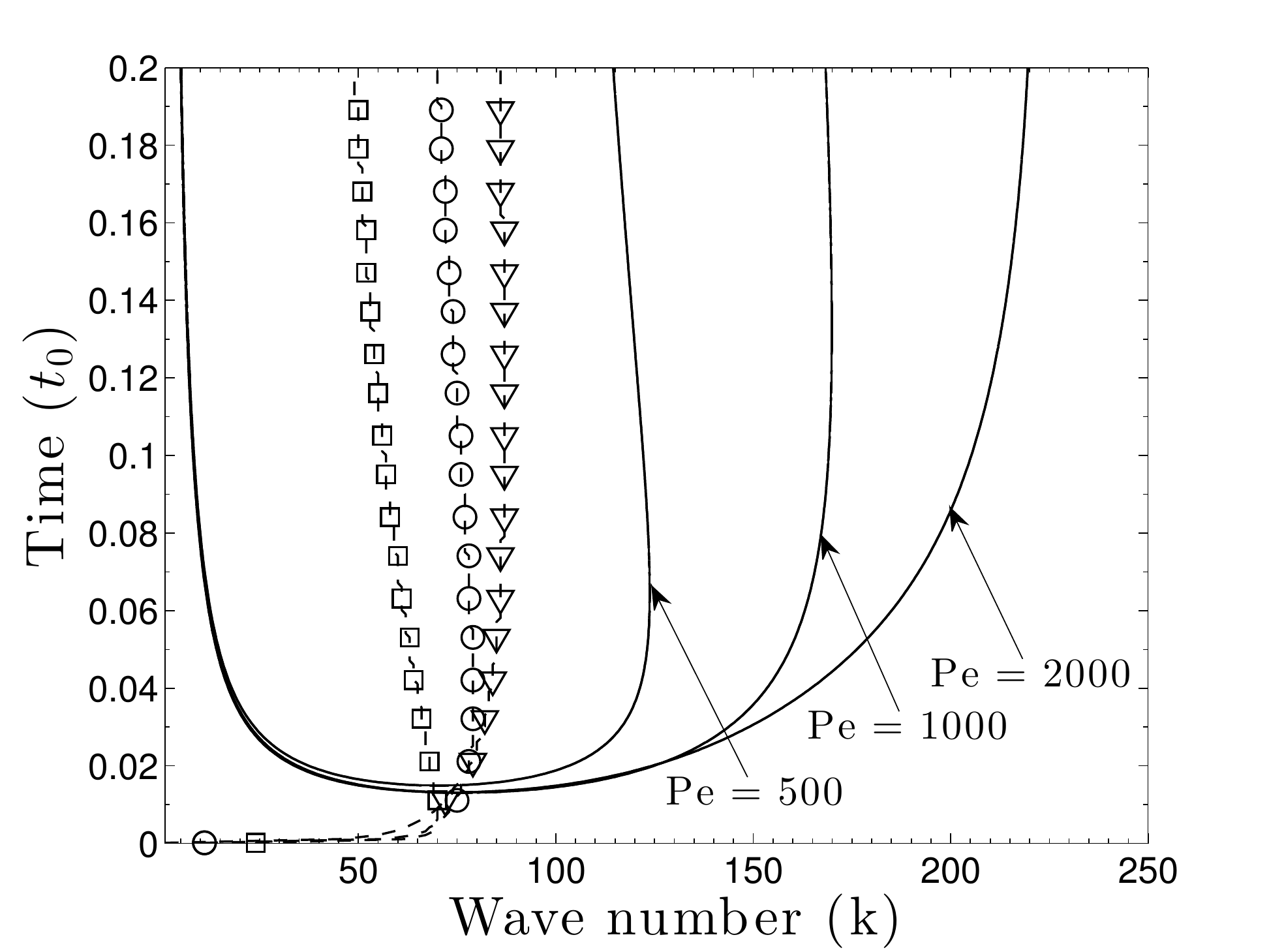}
\caption{Neutral stability curves ($\sigma^* = 0$) for different Pe values with $R = 3, \epsilon = 1$ and $\delta = -10^{-5}$. Lines with markers correspond the locus of the $k_{max}$ at successive times: Pe $= 500 (\Box)$, Pe $= 1000 (\bigcirc)$, Pe $= 2000 (\bigtriangledown)$.}\label{fig:KSneutral1}
\end{figure}

Fig. \ref{fig:KSneutral1} shows the neutral stability curves with $R = 3, \epsilon = 1, \delta = -10^{-5}$ for different values of Pe = $500, 1000, 2000$. Similar to the case of modeling without the Korteweg stresses, the region of instability increases for increasing Pe. However, few significant changes are observed, i.e. $k_{max}$ do not differ much with Pe unlike the case of $\delta = 0$, at least for Pe of same order. And the $k_{max}$ values, for different Pe, before the onset of fingering are smaller than those after the onset of instability, which were opposite in the absence of these stresses. It further depicts that both the $k_{cric}$ and $t_{cric}$ are almost independent of the values of Pe. It can be observed that for a fixed value of Pe, the region of instability decreases and $t_{cric}$ increases with the increasing magnitude of the Korteweg stress constant $\delta$. We further analyze the Korteweg stress influence on the stability with larger Pe values. Under this circumstance, the region of instability and the critical time $t_{cric}$ are found to change rapidly as $\delta$ increases in magnitude. 

%figure17
\begin{figure}[ht]
(a) \hspace{8 cm} (b) \\
\centering
\includegraphics[width=3.2in, keepaspectratio=true, angle=0]{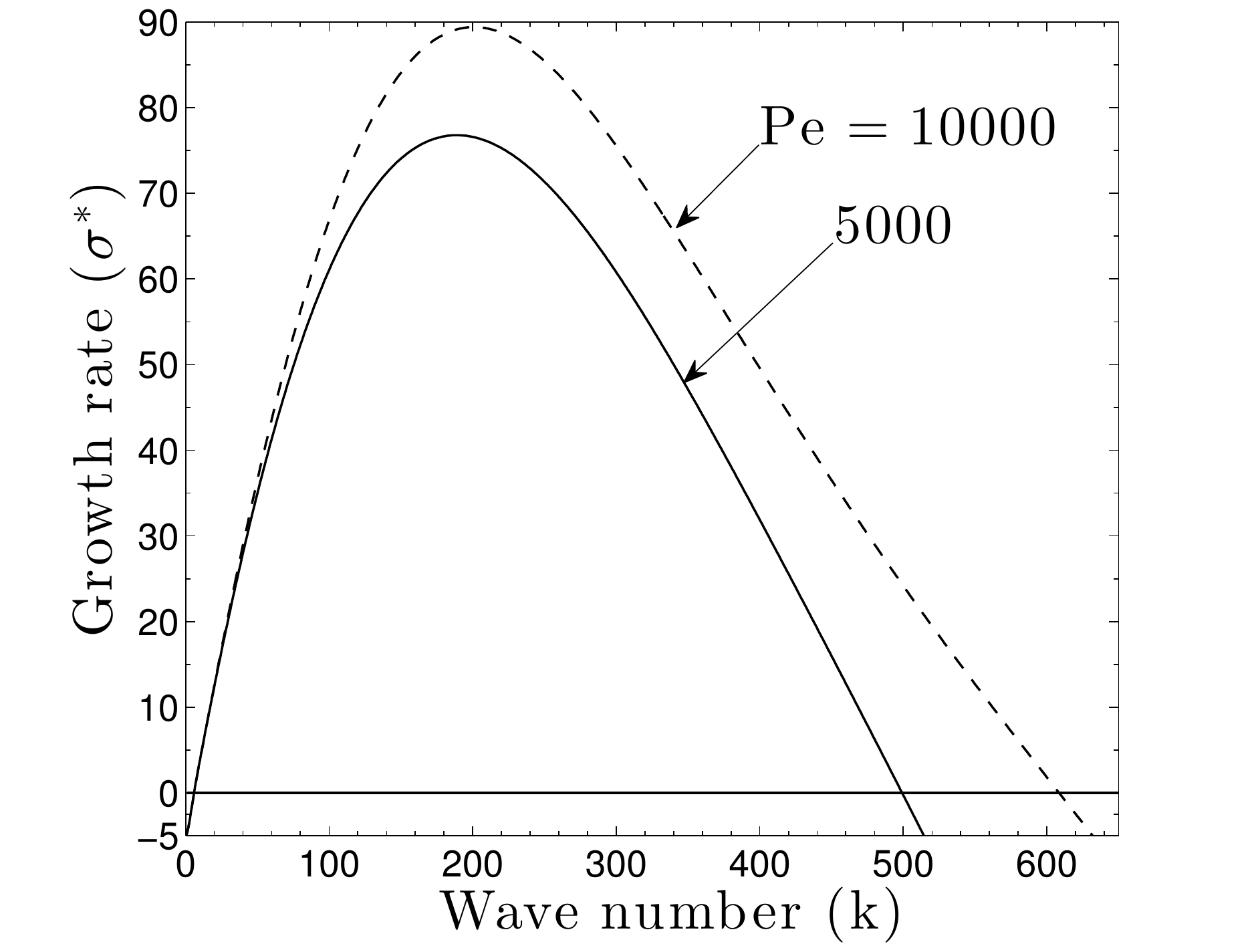}
\includegraphics[width=3.2in, keepaspectratio=true, angle=0]{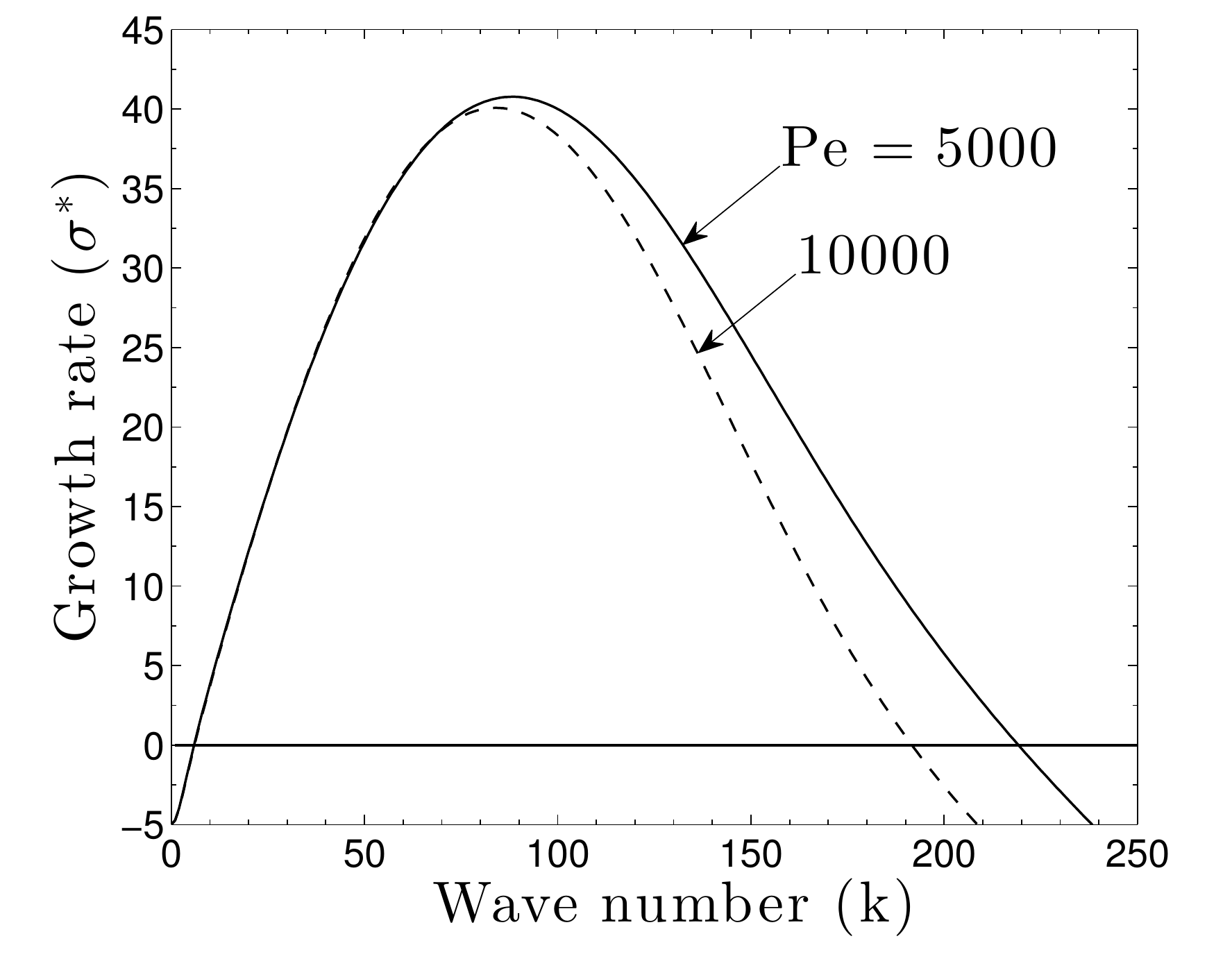}
\caption{Dispersion curves for $R = 3, \epsilon = 1$ at frozen diffusive time $t_0 = 0.1$ with (a) $\delta = -10^{-6}$, (b) $\delta = -10^{-5}$.}\label{fig:KSdispersion}
\end{figure}

A very important and surprising result has been observed while analyzing the dispersion curves, which are represented in Fig. \ref{fig:KSdispersion}. In particular, the dispersion curves presented in Fig. \ref{fig:KSdispersion}(a) for Pe $= 5000, 10000$ with $R = 3, \epsilon = 1, t_0 = 0.1$ and $\delta = -10^{-6}$ depicts the usual behavior, i.e. larger Pe corresponds to the larger growth rates. However, the existence of a reverse phenomenon is noticed for such choice of Pe when $\delta$ value has been changed from $-10^{-6}$ to $-10^{-5}$ (see Fig. \ref{fig:KSdispersion}(b)) keeping other parameters same as the case of Fig. \ref{fig:KSdispersion}(a). It is clearly observed that the higher Pe has less than or equal growth rate for all the wave numbers and this difference between the respective growth rates increase with $k$ as it crosses $k_{max}$. The possible reason is, for a very slow fluid dispersion, steepness of the concentration gradient gives rise to stronger effect of the Korteweg stresses, whose stabilization dominates the destabilization of slow dispersion. However, the occurrence of such reversal property of Pe effect on VF instability depends completely on the magnitudes of Pe, $\delta, R, \epsilon$ and $t_0$. Chen \textit{et al.} \cite{CWM} also reported similar observation through nonlinear simulations of the displacement of a miscible circular drop in a Hele-Shaw cell.

Here we look to investigate how the Korteweg stresses monitor the qualitative behavior of various critical parameters ($k_c, k_{cric}, t_{cric}$, etc.) as Pe value changes. It is observed that in the presence of the Korteweg stresses ($\delta = -10^{-5}$), for Pe $\leq \mathbf{O}(10^2)$ (in particular Pe $ \lesssim 500$) $k_c$ increases with Pe as $k_c \sim \text{Pe}^{0.7}$, whereas $k_{max}$ increases little slower, $k_{max} \sim \text{Pe}^{0.65}$. Again, for Pe $\geq \mathbf{O}(10^3)$ (specifically for Pe $\gtrsim 2000$) both $k_c$ and $k_{max}$ decrease, which corresponds the reversal behavior of Pe in the presence of the Korteweg stresses, as discussed earlier. Finally, the onset time of instability, $t_{cric}$, and the corresponding most unstable wave number,\cite{PM1} $k_{cric}$, are analyzed in the presence of the Korteweg stresses. It is observed that both $t_{cric}$ and $k_{cric}$ become almost constants, which were also clearly observed from the neutral stability curves (see Fig. \ref{fig:KSneutral1}). Nevertheless, in the nonlinear regime the scenario changes drastically. Based on the onset of nonlinear fingers ($t_{on}$) two different stability regimes are found in the presence of the Korteweg stresses. In particular for $R = 1, ~ \delta = -10^{-6}$ and Pe $\lesssim 2000, ~ t_{on}$ varies as Pe$^{-1.1}$; while for Pe $ \gtrsim 2000, ~ t_{on} \sim \text{Pe}^{-0.5}$. The first regime is dominated by the destabilization due to slow dispersion, whereas in the second regime stabilization of the Korteweg stresses is dominant. 

\section{Conclusions}\label{sec:C}
Influence of fluid dispersion on miscible VF has been analyzed by solving the governing equations in their dimensionless form, obtained using convective characteristic scales. It is observed that there exists a critical P\'{e}clet number, $\text{Pe}_c$, such that the rectilinear displacement becomes unstable if $\text{Pe} > \text{Pe}_c$. However, unlike the radial source flow (see Tan and Homsy \cite{TH3}), $\text{Pe}_c$ for the rectilinear displacement varies with $t_0$. 

It is further observed that with an increase in Pe the area of the unstable region increases rapidly. This signifies that slow dispersion leads to an intense unstable scenario. Two vital parameters $k_c$ and $k_{max}$ are found to increase with Pe; rate of increase is little higher for $k_c$ than $k_{max}$ for all values of frozen diffusive time $t_0$ and $R$. On the other hand, $k_{max}$ decreases faster than $k_c$ with $t_0$ independent of Pe and $R$. More number of fine scale fingers are formed when the dispersion becomes very slow (i.e. at higher Pe value) and the onset of fingering also happens to be earlier under this circumstance. 

In the absence of fingering instability $L_d$ changes as $\sqrt{t}$ and the proportionality constant $C_{diff}$ decreases with Pe at inverse square root (i.e. $C_{diff} \sim \text{Pe}^{-1/2}$). One of the most important findings of the present NLS is, unlike the other dimensionless formulation \cite{TH2}, the average wave number of the unstable modes increases while increasing the Pe values. This truly incorporates the influence of fluid dispersion and physically signifies that at a slow rate of fluid dispersion short wave disturbances also become unstable. 

It is observed that the onset time of nonlinear fingering $t_{on}$ decreases at a rate $10\%$ higher than that of $t_{cric}$, independent of $R$. In the presence of the Korteweg stresses two different stability regimes are found to exist having different power law dependence of $t_{on}$ on Pe. These findings are in resemblance with the LSA results in the presence of the Korteweg stresses in slow diffusive regime. Further, LSA with the Korteweg stresses shows a very interesting result about the most unstable wave number $k_{max}$. For a fixed nonzero value of $\delta$, $k_{max}$ remains almost unchanged with the change in Pe, unlike the case when $\delta = 0$. Besides that, $k_{max}$ in the stable regions are smaller than those of unstable regions in the presence of the Korteweg stresses which is also an opposite phenomenon in the absence of these stresses. With the inclusion of the Korteweg stress into the model, the variation of $k_c$ and $k_{max}$ with Pe becomes almost unpredictable that in general follow a particular relationship without such stresses. Anisotropic dispersion also plays a vital role on the onset of fingering instability. Smaller transverse dispersion helps in early onset both in the linear and nonlinear regimes. Usual experiments are performed for miscible VF at $U \sim 5$ mm/s to $50$ mm/s with the dispersion coefficient of $\mathbf{O}(10^{-3} ~\text{cm$^2$/s})$ in Hele-Shaw cells of average width $10^2~\text{mm}$, i.e. Pe $= \displaystyle\frac{UL_y}{D_x} \sim \mathbf{O}(10^3)$, which is an unstable configuration\cite{MRSMCD}. The present results can be useful for designing the experimental setup to obtain a stable displacement even for an unfavorable viscosity contrast between the miscible fluids by monitoring the displacement speed and the diffusion rate that generates the P\'{e}clet number below a critical value.

Influence of velocity induced dispersion and concentration dependent diffusion have been investigated in the literature both theoretically \cite{Zimmerman, Petitjeans, Riaz} as well as experimentally \cite{Petitjeans} in the context of miscible viscous fingering instability under different flow conditions. However, none of those studies explicitly mentioned about the influence of P\'{e}clet number Pe or Taylor dispersion P\'{e}clet number Pe$_{TD}$ on the onset of instability. In this paper we attempted to investigate such influences on the onset of fingering instability both qualitatively and quantitatively with constant dispersion and molecular diffusion. Inclusion of the velocity induced dispersion and concentration dependent molecular diffusion is beyond the scope of the present study and those will be addressed in our future studies.

\begin{acknowledgments}
The author M.M. gratefully acknowledges the financial support from the Department of Science and Technology (DST), Govt. of India. The author S.P. gratefully thanks the National Board for Higher Mathematics (NBHM), Department of Atomic Energy (DAE), Govt. of India for the Ph.D. fellowship.
\end{acknowledgments}

\appendix
\section{Analytic dispersion relation for step initial profile}\label{sec:ADR}
In this appendix we derive the analytic dispersion relation for a step initial concentration. However, the LSA in the self-similar $(\xi, y, t)$-coordinates is limited to the cases of non-zero frozen diffusive time, as $t = 0$ is a singular point for $\displaystyle\xi = \frac{x}{\sqrt{t}}$. Hence, an analytic expression for the dispersion relation at initial time has been derived in $(x, y, t)$-coordinate systems. This is followed by determining the dependence of growth rate, most dangerous wave number and the cutoff wave number on Pe at $t_0 = 0$. Following Tan and Homsy \cite{TH1}, application of QSSA method yields the following fourth order ordinary differential equation in the $(x, y, t)$-coordinate system,
\begin{equation}\label{eq:A1}
\left\{\frac{1}{\text{Pe}}\left(\frac{d^2}{dx^2} - k^2\right) - \sigma\right\}\left\{\frac{d^2}{dx^2} + R \frac{dc_0}{dx}\frac{d}{dx} - k^2\right\}\phi = Rk^2\frac{dc_0}{dx}\phi,
\end{equation}
associated with the boundary conditions $\phi \to 0$ and $\mathcal{L}\phi \to 0$ as $|x| \to \infty$. Here the differential operator $\mathcal{L}$ is defined as, $\mathcal{L} = R^{-1}k^{-2}\left(\displaystyle\frac{d^2}{dx^2} + R \frac{dc_0}{dx}\frac{d}{dx} - k^2\right)$; and the instantaneous growth rate of the perturbations, $\sigma$, is the eigenvelue of Eq. \eqref{eq:A1}. Taking the initial condition for the concentration as step profile we have $\displaystyle\frac{dc_0}{dx} = \delta(x)$ at $t_0 = 0$, where $\delta(x)$ represents the Dirac-delta function. Thus, Eq. \eqref{eq:A1} boils down to,
\begin{eqnarray}\label{eq:A2}
\left(\frac{d^2}{dx^2} - l^2\right)\left(\frac{d^2}{dx^2} - k^2\right)\phi = 0, ~~~~~~ \text{at} ~~ x \neq 0, 
\end{eqnarray}
where $l^2 = \sigma \text{Pe} + k^2$. Solving Eq. \eqref{eq:A2} associated with the appropriate boundary conditions the perturbed velocity $\phi$ and the corresponding concentration perturbation $\psi$ are obtained as, 
\[
\phi = \left\{\begin{matrix}
A_1e^{lx} + B_1e^{kx}, & x < 0 \\
A_2e^{-lx} + B_2e^{-kx}, & x > 0,
\end{matrix}
\right.
~~~~
\psi = \left\{\begin{matrix}
\displaystyle\frac{A_1(l^2 - k^2)}{Rk^2}e^{lx}, & x<0 \\
\displaystyle\frac{A_2(l^2 - k^2)}{Rk^2}e^{-lx}, & x>0,
\end{matrix}
\right.
\]
respectively. The arbitrary constants $A_1, A_2, B_1, B_2$ to be determined from the matching conditions,
\begin{eqnarray}
\label{eq:A3}
& & \phi(0^+) = \phi(0^-) \\
\label{eq:A4}
& & \psi(0^+) = \psi(0^-) \\
\label{eq:A5}
& & \alpha\frac{d\phi(0^+)}{dx} = \frac{d\phi(0^-)}{dx},
\end{eqnarray}
which correspond to the velocity continuity, concentration continuity and continuity of pressure, respectively, at the fluid-fluid interface.

Integrating Eq. \eqref{eq:A1} from $0^-$ to $0^+$ we get,
\begin{equation}\label{eq:A6}
\int_{0^-}^{0^+}\left(\frac{d^2}{dx^2} - l^2\right)\left(\frac{d^2}{dx^2} + R\delta(x)\frac{d}{dx} - k^2\right)\phi ~\text{d}x = \int_{0^-}^{0^+}\text{Pe}Rk^2\delta(x)\phi~\text{d}x.
\end{equation}
Using the properties of Dirac-delta function and performing some cumbersome algebra, dispersion relation can be determined from the following quadratic equation in $\sigma$Pe,
\begin{eqnarray}\label{eq:A7}
4(\sigma\text{Pe})^2 + 4k(k - R\text{Pe})\sigma\text{Pe} + k^2R\text{Pe}(R\text{Pe} - 4k) = 0.
\end{eqnarray}
Solving Eq. \eqref{eq:A7} for the instantaneous growth constant,
\begin{eqnarray}
\sigma\text{Pe} & = & \frac{-4k(k - R\text{Pe}) - \sqrt{16k^2(k - R\text{Pe})^2 - 16k^2R\text{Pe}(R\text{Pe} - 4k)}}{8} \nonumber \\
\label{eq:A8}
\Rightarrow \sigma & = & \frac{k}{2\text{Pe}}\left((R\text{Pe} - k) - \sqrt{k^2 + 2R\text{Pe}k}\right). 
\end{eqnarray}
(Other root of Eq. \eqref{eq:A7} corresponds a unphysical situation when the perturbation of the smallest wavelength has the largest growth rate.) We can retrieve the Eq. (42) of Tan and Homsy \cite{TH1} by taking $\text{Pe} = 1$ in Eq. \eqref{eq:A8}. Now, we look to investigate the dependence of $\sigma$ on \text{Pe}. Differentiation of Eq. \eqref{eq:A8} with respect to Pe gives,
\begin{eqnarray}\label{eq:A9}
\frac{d\sigma}{d\text{Pe}} = \frac{k}{2\text{Pe}^2}\left\{1 + \frac{k^2 + R\text{Pe}k}{\sqrt{k^2 + 2R\text{Pe}k}}\right\} \geq \left(\frac{k}{\text{Pe}}\right)^2 > 0, 
\end{eqnarray}
i.e. $\sigma$ is an increasing function of Pe.


\begin{thebibliography}{99}

\bibitem{H} G. M. Homsy, ``Viscous fingering in porous media," Annu. Rev. Fluid Mech. 19, 271-311 (1987).

\bibitem{DBM} A. De Wit, Y. Bertho, and M. Martin, ``Viscous fingering of miscible slices," Phys. Fluids 17, 054114 (2005).

\bibitem{MMD} M. Mishra, M. Martin, and A. De Wit, ``Differences in miscible viscous fingering of finite width slices with positive or negative log mobility ratio," Phys. Rev. E 78, 066306 (2008).
 
\bibitem{RDM} G. Rousseaux, A. De Wit, and M. Martin, ``Viscous fingering in packed chromatographic columns: linear stability analysis," J. Chromatogr. A 1149, 254-273 (2007).

\bibitem{MY} G. Morra, and D. A. Yuen, ``Role of Korteweg stresses in goedynamics," Geophys. Res. Lett. 35, L07304 (2008). 

\bibitem{ST} P. G. Saffman, and G. I. Taylor, ``The penetration of a fluid into porous medium or Hele-Shaw cell containing a more viscous liquid," Proc. R Soc. London Ser. A 245, 312 - 329 (1958).

\bibitem{HL} S. Hill, ``Channelling in packed columns," Chem. Eng. Sci. 1, 247-253 (1952).

\bibitem{TH3} C. T. Tan, and G. M. Homsy, ``Stability of miscible displacements in porous media: Radial source flow", Phys. Fluids 30, 1239 (1987).

\bibitem{YY} Z. Yang, and Y. C. Yortsos, ``Asymptotic solutions of miscible displacements in geometries of large aspect ratio," Phys. Fluids 9, 286 (1997).

\bibitem{TH1} C. T. Tan, and G. M. Homsy, ``Stability of miscible displacements in porous media: Rectilinear flow," Phys. Fluids 29, 3549-3556 (1986).

\bibitem{KC} M. C. Kim, and C. K. Choi, ``The stability of miscible displacement in porous media: Nonmonotonic viscosity profiles," Phys. Fluids 23, 084105 (2011).

\bibitem{PM1} S. Pramanik, and M. Mishra, ``Linear stability analysis of Korteweg stresses effect on miscible viscous fingering in porous media," Phys. Fluids, 25, 074104 (2013). 

\bibitem{TH2} C. T. Tan, and G. M. Homsy, ``Simulation of nonlinear viscous fingering in miscible displacement," Phys. Fluids 31, 1330 (1988).

\bibitem{ZH} W. B. Zimmerman, and G. M. Homsy, ``Nonlinear viscous fingering in miscible displacement with anisotropic dispersion," Phys. Fluids A 3, 1859 (1991).

\bibitem{CWM} C. Y. Chen, L. Wang, and E. Meiburg, ``Miscible droplets in a porous medium and the effects of Korteweg stresses," Phys. Fluids 13, 2447-2456 (2001).

\bibitem{J} D. D. Joseph, ``. Fluid dynamics of two miscible liquids with diffusion and gradient stesses," Eur. J. Mech., B/Fluids. 9(6), 565-596 (1990).

\bibitem{MFMG_2007} J. A. Pojman, N. Bessonov, V. Volpert, and M. S. Paley, ``Miscible fluids in microgravity (MFMG): A zero-upmass investigation on the international space station," Microgravity sci. technol. XIX-1, 33-41 (2007).

\bibitem{K} D. Korteweg, ``Sur la forme que prennet les \'{e}quations du mouvement des fluides si l'on tient compte des forces capillaires caus\'{e}es par des variations de densit\'{e}," Arch. N\'{e}erl Sci. Ex. Nat., Series II, 6, 1-24 (1901).

\bibitem{Fernandez2001} J. Fernandez, P. Kurowski, L. Limat, and P. Petitjeans, ``Wavelength selection of fingering instability inside Hele-Shaw cells," Phys. Fluids 13, 3120-3125 (2001). 

\bibitem{NB} D. A. Nield, and A. Bejan, \textit{Convection in porous media}, Springer, p. 15, 1992.

\bibitem{MH} O. Manickam, and G. M. Homsy, ``Simulation of viscous fingering in miscible displacments with nonmonotonic viscosity profiles," Phys. Fluids 6, 95-107 (1994).

\bibitem{Polymer_Monomer_2005} N. Bessonov, V. A. Volpert, J. A. Pojman, and B.D. Zoltowski, ``Numerical Simulations of Convection Induced by Korteweg Stresses in Miscible Polymer-Monomer Systems," Microgravity Sci. Tech., 17(1), 8-12 (2005). 

\bibitem{Pojman_Langmuir_2006} J. A. Pojman, C. Whitmore, M. L. T. Liveri, R. Lombardo, J. Marszalek, R. Parker, and B. Zoltowski, ``Evidence for the Existence of an Effective Interfacial Tension between Miscible Fluids: Isobutyric Acid-Water and 1-Butanol-Water in a Spinning-Drop Tensiometer," Langmuir, 22, 2569-2577 (2006).

\bibitem{Zoltowski_Langmuir_2007} B. Zoltowski, Y. Chekanov, J. Masere, J.A. Pojman, and V. Volpert, ``Evidence for the Existence of an Effective Interfacial Tension between Miscible Fluids. 2. Dodecyl Acrylate-Poly(Dodecyl Acrylate) in a Spinning Drop Tensiometer," Langmuir, 23, 5522-5531 (2007). 

\bibitem{MRSMCD} R. Maes, G. Rousseaux, B. Scheid, M. Mishra, P. Colinet, and A. De Wit, ``Experimental study of dispersion and miscible viscous fingering of initially circular sample in Hele-Shaw cells," Phys. Fluids 22, 123104 (2010). 

\bibitem{Fernandez} E. J. Fernandez, C. A. Grotegut, G. W. Braun, K. J. Kirschner, J. R. Staudaher, M. L. Dickson, and V. L. Fernandez, ``The effects of permeability heterogeneity on miscible viscous fingering: A three dimensional magnetic resonance imaging analysis," Phys. Fluids 7, 468 (1995). 

\bibitem{Hu_ZAMP_1992} H. Hu and D. D. Joseph, ``Miscible displacement in Hele-Shaw cell,” Z. Angew. Math. Phys. 43, 626 (1992). 

\bibitem{Sugii} Y. Sugii, K. Okamoto, A. Hibara, M. Tokeshi, T. Kitamori, ``Effect of Korteweg
stress in miscible two-layer flow in a microfluidic device," Journal of Visualization
8 (2), 117–124 (2005).

\bibitem{Zimmerman}W.B. Zimmerman and G.M. Homsy, ``Viscous fingering in miscible displacements: Unification of effects of viscosity contrast, anisotropic dispersion, and velocity dependence of dispersion on nonlinear finger propagation" Phys. Fluids A 4, 2348 (1992).

\bibitem{Petitjeans} P. Petitjeans, C.-Y. Chen, E. Meiburg, and T. Maxworthy, ``Miscible quarter five-spot displacements in a Hele-Shaw cell and the role of flow-induced dispersion," Phys. Fluids 11 ,1705 (1999).

\bibitem{Riaz} A. Riaz, C. Pankiewitz, and E. Meiburg ``Linear Stability of Radial Displacements in Porous Media: Influence of Velocity-Induced Dispersion and Concentration-Dependent Diffusion," Phys. Fluids 16, 3592 (2004).

\end{thebibliography}
\end{document}